\documentclass[aps,prd,twocolumn,superscriptaddress,showpacs,floatfix]{revtex4-1}
\usepackage[utf8]{inputenc}
\usepackage{amsmath}
\usepackage{amsfonts}
\usepackage{amssymb}
\usepackage[final]{graphicx}
\usepackage[caption = false]{subfig}
\usepackage{epstopdf}
\usepackage{tabularx}
\usepackage{mathtools}
\usepackage[mathscr]{eucal}
\usepackage{xcolor}
\usepackage{epsfig}
\usepackage{float}
\usepackage{soul}
\usepackage{lineno}

\newcommand{\sarkar}[1]{\textcolor{blue}{#1}}

\begin{document}
\title{Defect production and quench dynamics in the three-dimensional Kitaev model}
\author{Subhajit Sarkar}
\email{sbhjt72@gmail.com, subhajit@post.bgu.ac.il}
\affiliation{Institute of Physics, P.O.: Sainik School, Bhubaneswar 751005, Odisha, India.}
\affiliation{Homi Bhabha National Institute, Mumbai - 400 094, Maharashtra, India}
\affiliation{Department of Chemistry, Ben Gurion University of the Negev, Beer Sheva 8410501, Israel.}
\author{Dibyendu Rana}
\email{dibyendurana@iopb.res.in}
\affiliation{Institute of Physics, P.O.: Sainik School, Bhubaneswar 751005, Odisha, India.}
\affiliation{Homi Bhabha National Institute, Mumbai - 400 094, Maharashtra, India}
\author{Saptarshi Mandal}
\email{saptarshi@iopb.res.in}
\affiliation{Institute of Physics, P.O.: Sainik School, Bhubaneswar 751005, Odisha, India.}
\affiliation{Homi Bhabha National Institute, Mumbai - 400 094, Maharashtra, India}
\date{\today}
\begin{abstract}\label{abs}
We study the quench dynamics of the three-dimensional Kitaev (spin) model under a linear drive using both exact numerical calculations and analytical ``independent crossing approximation". Unlike the two-dimensional Kitaev model, the three-dimensional Kitaev model reduces to a multilevel Landau-Zener problem for each momentum. We show that for the slow quench, the defect density is proportional to the quench rate $1/\tau$. We find that the zeros of the relevant coupling between the levels determine the non-adiabatic condition for the production of defects. The contour on which the energy spectrum becomes gapless does not play an active role. The asymptotic behavior of the defect density crucially depends on the way the system reaches the non-adiabatic regime during the quenching process. We analytically show that defect correlation varies as $\tau^{-1} e^{-A/\tau}$, where $A$ is a constant independent of $\tau$. For the slow quench, the qualitative dependence of the   entropy (produced during the quenching process) on the quench time is the same as that of the defect correlation, indicating a close connection between the defect correlation and the entropy content of the final state. Possible experimental realization of such quench dynamics is also described briefly.

\end{abstract}
\maketitle
\section{Introduction}
 Properties of physical systems near a quantum phase transition are determined by the change of the symmetry of the ground state. In particular, at low temperatures, a quantum phase transition is observed when the inherent quantum fluctuations win over the thermal fluctuations and determine the ground state properties \cite{qpt1, qpt2, qpt3}. In this respect, it is pertinent to ask what happens at absolute zero temperature, when the parameters of a given Hamiltonian (or physical system) are driven or varied in time and the system is taken across the different phases. Recently such driven quantum systems have attracted a lot of interest \cite{Pol1, Dzi}. On a more fundamental note, it is appropriate to ask if the quantum system always remains in its instantaneous ground state or fail to do so due to the diverging length and time scales, when taken across the critical region very slowly (by varying the relevant system parameters linearly with time) \cite{qpt1, qpt2, qpt3, Pol1, Dzi, MH, Son}. Consequently, as the quantum critical point (QCP) is approached the system gradually enters into a non-adiabatic regime by failing to remain in its instantaneous ground state. As a result, excitations are created which are also called the defects \cite{Pol1, Dzi}.

 In a generic second-order phase transition, the defect density, produced during quench across the QCP, scales with the quench time as $\tau^{-\frac{d \nu}{(z \nu+1)}}$, $d $ being the dimension of the system, $\nu$ and $z$ are the correlation length exponent and the dynamical critical exponent respectively \cite{Kib1, Kib2, Zur1, Zur2, Zur3, Dzi, Pol1, Dam}. It is found explicitly that in low dimensions the response of the system to the slow changes of the Hamiltonian parameters could be non-analytic and non-adiabatic \cite{Pol2}. It has been shown that for a sufficiently slow quench at a rate $1/\tau$ $(<<1)$,
 the density of the above mentioned defect states scales with the quench time as, $n_d \sim \tau^{-(\frac{m\nu}{z\nu +1})}$, when the process of quenching takes the system across a $d-m$ dimensional gapless (critical) hyper-surface instead of a QCP \cite{Pol3, Dzi, KS1, KS2}. Such a scaling behavior has been studied in the two-dimensional (2D) Kitaev model as well as in several other exactly solvable models \cite{Dzi, KS1, KS2}.

 The Kitaev model is a rare example of a fully anisotropic spin model which is exactly solvable \cite{Kit}. The presence of Kitaev-like interactions has been found in a class of  materials, viz., `iridate' systems \cite{Jac, YS1, YS2, Plum}. Several aspects of the Kitaev model have been explored theoretically such as, topological order, fractionalization of spins, quantum spin liquid, entanglement generation, as well as in the context of quantum computation \cite{Nay, Bal, Her, And, MS, Wen, Lac, SM-GB,SM-GB-RS,SM-MM}. A three-dimensional (3D) version of the Kitaev model has been proposed and studied theoretically, and has been realized experimentally in layered `iridate' compounds \cite{SM, Oka, Mod, Taka}. The 3D Kitaev model contains four sub-lattices in an unit cell and its energy spectrum vanishes on a contour. Furthermore, it supports fermionic and bosonic non-local  excitations as well \cite{SM-NS}.
 
 Motivated by the above studies, we investigate the quench dynamics in the spin-1/2 3D Kitaev model on a hyper-honeycomb lattice. Due to the presence of a four-sublattice structure, the model, in its ground state flux configuration, exhibits a pair of negative energy and positive energy states for each momentum. This enables us to explore a multi-level Landau-Zener (L-Z) problem in the quench dynamics of the 3D Kitaev model. By filling up both the negative energy states, we end up in a six-level L-Z problem. On the other hand, if we start with the most negative energy state being filled, the model reduces to a four-level L-Z problem. Although such a multi-level L-Z problem (both the four-level \cite{Sin1, Sin2, Sin3} and beyond \cite{Tatsuya, Demkov}) has been studied using model Hamiltonians, however, most of them are either of mathematical interest or lack a direct connection to the real materials. The question of the exact solvability of multi-level L-Z problems has also attracted a lot of interest \cite{chernyak2020integrable, Chernyak_2018}. It is therefore intriguing to study the quench dynamics of the 3D Kitaev model (in the multi-level set-up) which is both exactly solvable and experimentally realizable in a class of materials mentioned earlier.
 \paragraph*{}
 In this paper, we investigate the defect density, defect correlation, and entropy in the final state of the quench dynamics in the 3D Kitaev model. We define the defect correlation as the two-point fermionic correlation in the defect state (time-evolved initial state) \cite{Pol2,KS1,KS2}. Furthermore, quenching pushes the system to make transitions to other energy levels via non-adiabatic points. 
 Moreover, we calculate the von-Neumann entropy of the final state \cite{nielsen_chuang_2010, Neumann,levitov-2006}.  Interestingly enough, we find that the 3D Kitaev model offers a solvable physical system where the semi-classical ``Independent Crossing Approximation(ICA)" can be applied to obtain an analytical form for the defect density, thereby enlarging our scope of understanding of the multi-level L-Z dynamics. Remarkably, we observe that in the multi-level L-Z dynamics of 3D Kitaev model the non-adiabatic condition between the relevant levels is not determined entirely by the spectrum of the model, rather by the coupling between the adiabatic levels. This might lead to a scaling of the defect density beyond the conventional two-level L-Z dynamics.
 We further investigate the defect correlation and obtain an analytical expression for it which agrees with the corresponding numerical results. The entropy is found to be generated due to the non-local correlation developed during the quench.
 \paragraph*{}
 The paper is organized as follows. In section \ref{mod}, we introduce the 3D Kitaev model as conceived in Ref \cite{SM}. We explain in detail the methods to transform the Hamiltonian in momentum space into the one that is suitable for the quench dynamics study. Here we elaborate our scheme to reduce the 3D Kitaev model to a multi-level L-Z problem (both four and six levels) and describe the numerical scheme to solve the time dependent Schrodinger equations. In Sec. \ref{ica}, we present the analytical formula for the defect density within the semi-classical ``Independent Crossing Approximation (ICA)". Sections \ref{defect}, \ref{correl}, and \ref{ent} describe the results corresponding to the defect density, defect correlation and the entropy of the final state respectively, obtained numerically using the method outlined in \ref{num-scheme}, and analytically as well using ICA (whenever applicable). We finally conclude and discuss our results in section \ref{conc}. In the Appendices \ref{ta} and \ref{ta1} we briefly discuss the issue of dynamical phases in the transition probabilities associated with semi-classical trajectories and justify the ICA. The other Appendices describe various relevant details related to our calculations.
\section{The Model and Formalism}\label{mod}
\begin{figure}[h]
\includegraphics[scale=0.25]{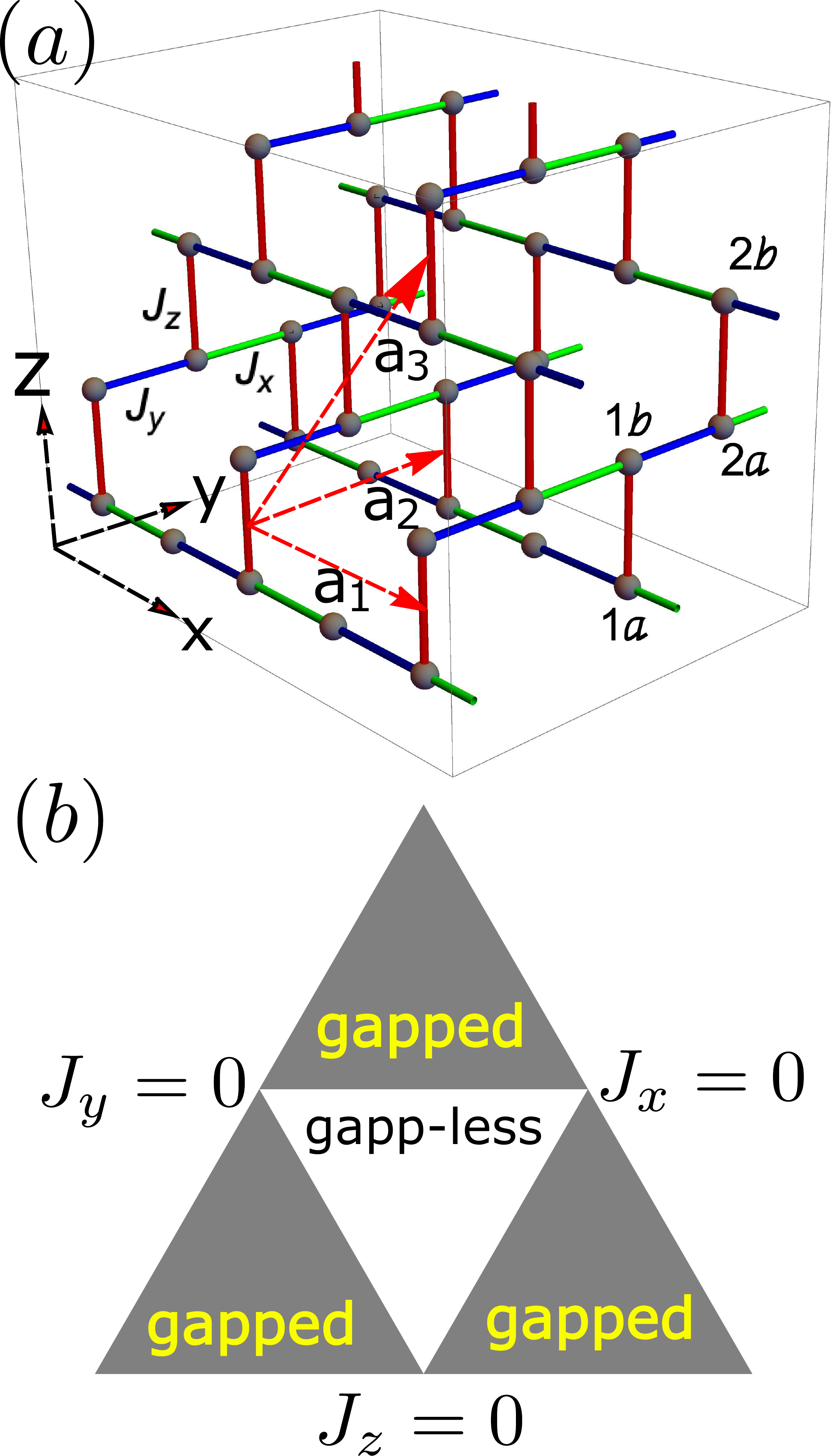}
\caption{\textbf{Three-Dimensional Kitaev model, the lattice and the phase diagram}: (a) A part of the three dimensional hyper-honeycomb lattice on which 3D Kitaev model is defined. The unit cell consist of two vertical $z$-type bonds connected by a $y$-type bond along $y$-direction. $(1a ,~ 1b)$ corresponds to he dimer $\mu = 1$ with $(a,~b)$ being the sub-lattice index. Similarly, $(2a ,~ 2b)$ corresponds to he dimer $\mu = 2$. (b) The phase diagram of the 3D Kitaev model in the parameter space $(J_x, J_y, J_z)$. The three sides of the triangle correspond various limits as shown. The gray shaded region is the gapped phase and the middle region is the gapless phase.}
\label{3d-phs-d}
\end{figure}

The three-dimensional (3D) Kitaev model is defined on a hyper-honeycomb lattice with lattice coordination number `3' (see Fig.~\ref{3d-phs-d} (a)). 
Each spin interacts with its three nearest neighbors through three different links, viz., $x$-link, $y$-link, and $z$-link which only exhibit interaction between the corresponding spin components \cite{SM}. The spin Hamiltonian on the hyper-honeycomb lattice has the following form,
\begin{equation}
H = -J_x \sum_{x-link} \sigma_{i}^{x} \sigma_{j}^{x} -J_y \sum_{y-link} \sigma_{i}^{y} \sigma_{j}^{y} -J_z \sum_{z-link} \sigma_{i}^{z} \sigma_{j}^{z}.
\label{ham-init}
\end{equation}
In the above equation, `$i,j$' represents a pair of nearest neighbor spins and the sum is over each type of bonds. The unit cell contains four sites with the basis vectors given by, $\mathbf{a}_1 = 2 \hat{\mathbf{x}}$, $\mathbf{a}_2 = 2 \hat{\mathbf{y}}$, and $\mathbf{a}_3 = (\hat{\mathbf{x}}+ \hat{\mathbf{y}} +2 \hat{\mathbf{z}})$ \cite{SM}.

Following Kitaev's original prescription for solving the 2D Kitaev model, the above Hamiltonian can be fermionized to reduce to a non-interacting Majorana Fermion hopping problem in the presence of conserved $\mathbb{Z}_2$ gauge fields defined on every links/bonds. These $\mathbb{Z}_2$ gauge fields can take values $\pm 1$. For each configuration of these $\mathbb{Z}_2$ gauge fields, one obtains a fermionic spectrum. It is found that~\cite{SM}, the ground state sector corresponds to the configuration where the product of $\mathbb{Z}_2$ gauge fields over  an elementary plaquette is one. We chose the $\mathbb{Z}_2$ gauge field to be equal to `$+1$' for each bond,
and this enables us to perform a Fourier transform and work in the momentum space.  In the ground state sector (where all $\mathbb{Z}_2$ gauge fields are uniformly `$+1$') the effective Majorana fermion Hamiltonian  is given by,
\begin{eqnarray}\label{kitaev_r}
&&H = i \sum_{\mathbf{r}} \Big[ \sum^{1,2}_j J_z c_{ja} (\mathbf{r})c_{jb} (\mathbf{r}) +J_x c_{1a} (\mathbf{r})c_{2b} (\mathbf{r_3})+  \nonumber \\ &&J_y c_{1a} (\mathbf{r})c_{2b} (\mathbf{r_{13}}) +  J_y c_{1b} (\mathbf{r}) c_{2a} (\mathbf{r_2}) + J_x c_{1b} (\mathbf{r})c_{2a}(\mathbf{r})   \Big],~~
\end{eqnarray}
 where $\mathbf{r}$ is a position vector of a lattice point \cite{SM} and $\mathbf{r}_3=\mathbf{r}-\mathbf{a}_3, \mathbf{r}_2=\mathbf{r}-\mathbf{a}_2$ and $\mathbf{r}_{13}=\mathbf{r} + \mathbf{a}_1-\mathbf{a}_3$.  Owing to the bipartite nature of the lattice, in the above Hamiltonian, we  conveniently introduce two indices to label the sites within a unit cell; $\mu = 1, 2$, denote the dimer to which a site belongs and $\alpha= a, b$, denote the sub-lattice indices \cite{SM}. The spectrum of the Hamiltonian in \eqref{kitaev_r} is found to be gapless in the middle white region as shown in Fig.~\ref{3d-phs-d} (b). 
 
 In our quench study, we take $J_z$ to be increasing linearly with time $t$, i.e., $J_z (t)=Jt/\tau$ at a rate $1/\tau$. To implement the quench study in the  L-Z set-up, we rewrite the Hamiltonian in \eqref{kitaev_r} by  introducing  the complex fermions $\psi$ and $\phi$ by regrouping the two Majorana fermions at two z-bonds of the unit cell as $\psi_i = c_{i,1a} + i c_{i,1b},~~~\phi_i = c_{i,2a} + i c_{i, 2b}$. A subsequent Fourier transform of \eqref{kitaev_r} leads us to, $H = \sum_{k}^{'} \Psi_{k}^{\dagger} h(k) \Psi_{k}$, where `$k$' belongs to half Brillouin  zone (HBZ, denoted by the primed summation). Here,
\begin{eqnarray}
\label{hkmat}
h(k) &=& \begin{pmatrix}
2 J_{z} & -\Delta^{*}_{2k} & 0 & -\Delta^{*}_{1k} \\
-\Delta_{2k} & 2  J_z & \Delta_{1k} & 0 \\
0 & \Delta^{*}_{1k} & -2J_z & \Delta^{*}_{2k}\\
-\Delta_{1k} & 0 & \Delta_{2k} & -2 J_z
\end{pmatrix}, \; 
\end{eqnarray}
and $
\Psi_{k} = \begin{pmatrix}
\psi_{k}^{\dagger} & \phi_{k}^{\dagger} & \psi_{-k} & \phi_{-k}
\end{pmatrix}^{\dagger}$,
where $\Delta_{1k} = e^{i\mathbf{k\cdot a_3}}\delta_{1k} +\delta_{2k}$, and $\Delta_{2k} = e^{-i\mathbf{k\cdot a_3}}\delta_{1k}^{*} - \delta_{2k}^{*}$, and $\delta_{j,k}= J_x + J_y e^{i (-1)^j k_j}$ with $j=1,2$. 
Diagonalizing the upper-left and lower-right  $2 \times 2$ blocks we get a couple of eigenvalues, $\epsilon_{1k} = 2J_z + |\Delta_{2k}|$ and $\epsilon_{2k} = 2J_z - |\Delta_{2k}|$ corresponding to the upper-left block, and another couple  $-\epsilon_{1k}$ and $-\epsilon_{2k}$ corresponding to the lower-right block respectively. Using the unitary transformation corresponding to the above mentioned diagonalization (see Appendix \ref{tb} for explicit form), we re-write  \eqref{hkmat} as, 
\begin{eqnarray}\label{4LZ}
\tilde{h}(k) &=& \begin{pmatrix}
\epsilon_{1k} & 0 & g_k & -\gamma_k \\
0 & \epsilon_{2k} & \gamma_k & -g_k  \\
-g_k & \gamma_k & -\epsilon_{1k} & 0\\
-\gamma_k & g_k & 0 & -\epsilon_{2k}
\end{pmatrix}.
\end{eqnarray}
where $g_k = -i|\Delta_{1k}| \sin \theta_{12k}$ and $\gamma_k = -|\Delta_{1k}| \cos \theta_{12k}$ with $e^{i\theta_{12 k}} = \frac{\Delta_{1k}\Delta_{2k}^{*}}{|\Delta_{1k}||\Delta_{2k}|}$, and $\epsilon_{1k} = 2J_z + |\Delta_{2k}|$ and $\epsilon_{2k} = 2J_z - |\Delta_{2k}|$.
We denote $\alpha_{i,k}$($i=0,1,2,3$) as the annihilation operators corresponding to the
four states in the rotated basis. The relations between the $\alpha_{ik}$ to the $\psi_{k}$ and $\phi_k$ are given in Appendix \ref{tb}. We apply the quench protocol
by keeping $J$ fixed for a set of values of $\alpha =\tan^{-1}\frac{J_y}{J_x}$ quantifying the ratio of $J_{x}$ and $J_{y}$ .

The Hamiltonian corresponding to \eqref{4LZ} offers two ways of studying the quench dynamics depending on if we start with the initial state as the lowest negative energy state being occupied or start by filling up both the negative energy states. In the former case, we end up with a four-level L-Z problem, and in the latter, we find a six-level L-Z problem. It is worthwhile to emphasize that the ground state of the six-level problem belongs the true ground state of the spin-model at $t=-\infty$. Therefore, we denote the lowest energy state at $t=-\infty$ corresponding to the six-level problem as the ground state and the same corresponding to the four-level problem as the initial state. However, we shall henceforth use them synonymously and expect their meaning to be clear in the context. Below we discuss both of these two possibilities one by one. Furthermore, in connection to the original spin model corresponding to \eqref{ham-init}, this physically means that for the four-level L-Z problem, the initial state is the ferromagnetic state of the lower $z$-dimer of the unit cell. On the other hand, for the six-level problem, the initial state is a ferromagnetic state of both the $z$-dimers of the unit cell.
 
 \subsection{As a four-level problem}
\begin{figure}[h!]
\includegraphics[scale=0.195]{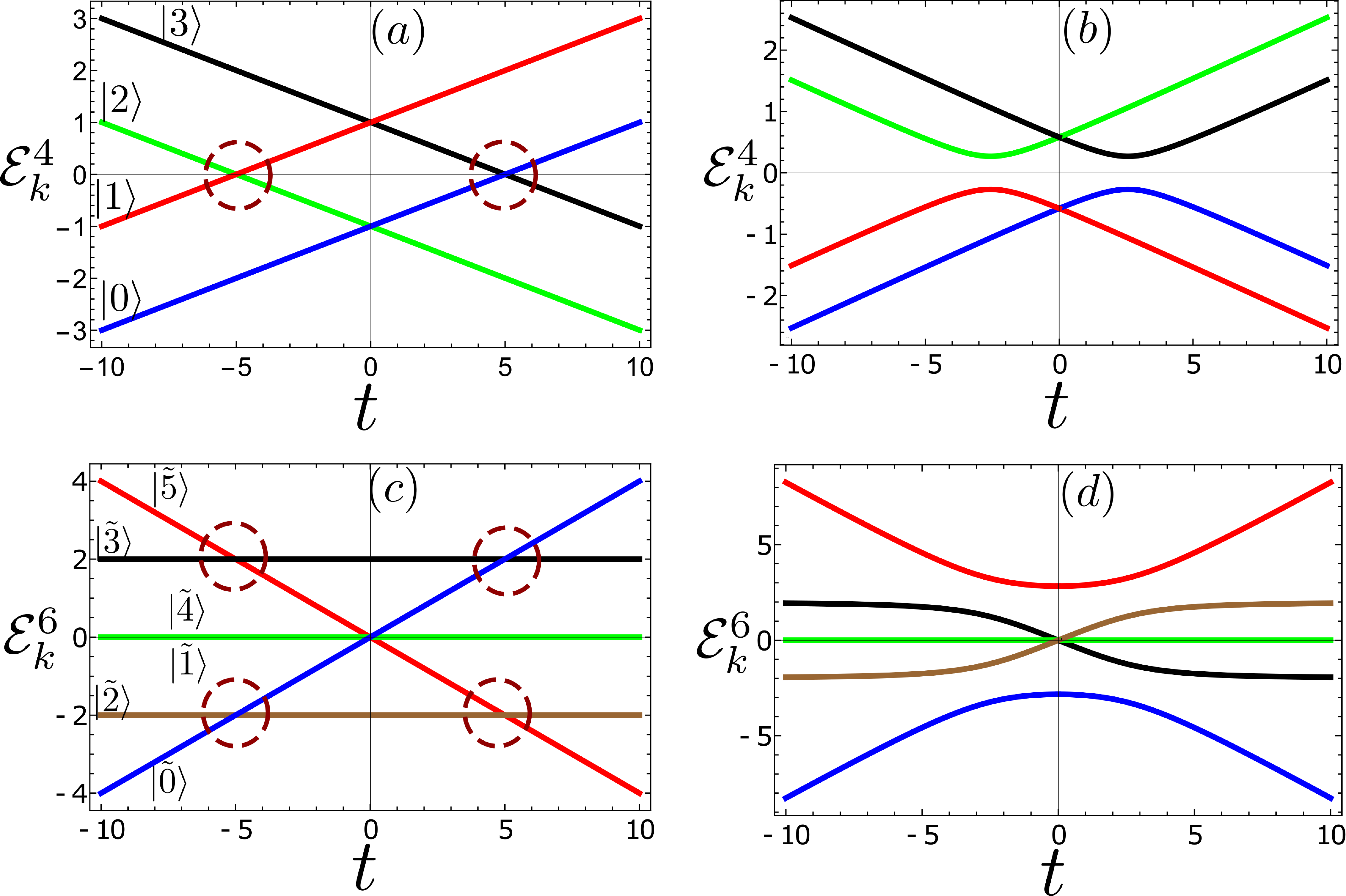}
\caption{\textbf{Energy-level diagrams:} (a) Diabatic energy levels corresponding to the Hamiltonian \eqref{4state-coupled} when only the diagonal elements are present. $|0\rangle$ is the initial state at $t=-\infty$ which become one of the excited states at $t=+\infty$, whereas $|2\rangle$ is the lowest energy state at $t=+\infty$ which has been one of the excited states at $t=-\infty$.  At $t\rightarrow - \infty$, the states $|0\rangle, |1\rangle , |2\rangle , \, \text{and} \, |3\rangle$ have eigenvalues $\epsilon_{2k} , \, \epsilon_{1k} , \, -\epsilon_{1k} , \,\text{and} \, -\epsilon_{2k}$ respectively. (b) Adiabatic levels corresponding to the same Hamiltonian.
(c) Diabatic energy levels corresponding to the six-level representation corresponding to \eqref{6state-coupled} when only the diagonal elements are present.  At $t\rightarrow - \infty$, the states $|\tilde{0}\rangle, |\tilde{1}\rangle , |\tilde{2}\rangle , |\tilde{3}\rangle,|\tilde{4}\rangle, ~~ \text{and} ~~ |\tilde{5}\rangle$ have eigenvalues $4J_{3},~~ 0,~~ -2|\Delta_{2k}|,~~  2|\Delta_{2k}|,~~ 0,~~\text{and} ~~ -4J_{3}$ respectively. (d) Plot of the adiabatic levels for the six-state representation of the 3D Kitaev model. The plots are the representative of the avoided level crossing at a particular value of $k$ within the HBZ. It is worthwhile to note that there are only two and four avoided level crossings while there are four and six diabatic states in the four- and the six-level problems respectively. The red circles indicate the level crossings in the diabatic limit which eventually lead to avoided level crossings in the diabatic limits. The avoided crossing points are quite far apart, maintaining the applicability of the independent crossing approximation.}
\label{dia-adia-label}
\end{figure}

 In the limit of $t \rightarrow \pm \infty$, the dominating terms in \eqref{4LZ} are the diagonal ones where we retain a very small non-zero value of $|\Delta_{2k}|$, given the fact that $J_{x(y)}$ have infinitesimal but non-zero values. This helps us to avoid the obvious degeneracy (which otherwise originates from the fact that two of the diagonal elements of \eqref{4LZ} become equal) in the limit $t\rightarrow \pm \infty$, in the same spirit of removing the ground state degeneracy of a spin-$S$ Ferromagnetic Heisenberg model with an infinitesimally small symmetry breaking field. In the limit of $t\rightarrow - \infty$ the initial state is gapped, and the corresponding eigenvalue is given by, $\epsilon_{2k} = -2J_z - |\Delta_{2k}|$ with $|\Delta_{2k}|$ very small (but non-zero) so that the diabatic levels are well separated at $t\pm \infty$, as plotted in Fig.\ref{dia-adia-label}(a). According to our convention, at $t\rightarrow -\infty$, the eigenstates corresponding to $\epsilon_{2k}, \epsilon_{1k}, -\epsilon_{1k}$ and $-\epsilon_{2k}$ are denoted by  $|0 \rangle_k, | 1 \rangle_k, | 2 \rangle_k$, and $| 3 \rangle_k$ respectively. The diabatic ground state/initial state in this time limit is given by, $|G_{-\infty}\rangle = \begin{pmatrix} 1&0&0&0 \end{pmatrix}^{\dagger}$  whereas, in the limit $t\rightarrow + \infty$ the diabatic initial state is $|G^{'}_{+\infty}\rangle = \begin{pmatrix} 0&0&1&0 \end{pmatrix}^{\dagger}$. The corresponding adiabatic levels are plotted in Fig.\ref{dia-adia-label}(b). 
\paragraph*{}
The quench dynamics of various model systems have been studied in the past, and these, in essence, reduce to a two-level L-Z problem~\cite{Pol1,Pol2,KS1,KS2} which has been analytically examined. However, the system we are interested in turns into a multi-level L-Z model for each momentum $\mathbf{k}$, which limits the possibility of a complete analytical solution. In these circumstances, the numerical analysis of the evolution of the system under quench turns out to be the most convenient to determine the long time behavior of the system. We need to solve a set of coupled first-order differential equations corresponding to the time-evolution, which is given by, $i\frac{d}{dt} \Phi_{4,k}= \mathcal{U}_k \Phi_{4,k}$, where $\Phi^{\dagger}_{4,k}=(|0,t \rangle, |1,t \rangle, |2,t \rangle, |3,t \rangle)$, and the $4 \times 4$ matrix $\mathcal{U}_k$ is given below,
\begin{eqnarray}\label{4state-coupled}
\mathcal{U}_k=\left(\begin{array}{cccc}
\epsilon_{2k}&0& \gamma_k&-g_k \\
0& \epsilon_{1k}& g_k&-\gamma_k \\
\gamma_k&-g_k&-\epsilon_{1k}&0 \\
g_k& - \gamma_k&0& -\epsilon_{2k} \end{array}
\right),
\end{eqnarray}
 is the time evolution operator which is obtained from \eqref{4LZ} by an unitary transformation, $U_{k} = \left(\begin{array}{cc}
\sigma_{x}&0 \\
0& I_{2\times2} \end{array}
\right).$

\subsection{As a six-level problem}
 We now describe the quench dynamics
 when the initial state is constructed by filling up both the negative energy states corresponding to \eqref{4LZ} at $t \rightarrow - \infty$. The possibility of transition from the ground state to any state having two particles results into a six-level L-Z problem. To this end one needs to define the two particle sector of the Hilbert space and it is given in Appendix \ref{tb}. Time evolution is governed by the usual Schrodinger equation $i \frac{d}{dt} \Phi_{6,k}(t)= \tilde{\mathcal{U}}_k \Phi_{6,k}(t)$, where $\tilde{\mathcal{U}}_k$ is given by,
\begin{eqnarray}\label{6state-coupled}
\tilde{\mathcal{U}}_{k}= \left(  \begin{array}{cccccc}
4 J_3 & \gamma_k &  g_k & - g_k & - \gamma_k & 0 \\
\gamma_k & 0 & 0& 0& 0& - \gamma_k \\
+g_k & 0 & -2 |\Delta_{2k}|& 0 & 0 & - g_k \\
- g_k & 0 & 0 & 2|\Delta_{2k}| & 0 & +g_k \\
-\gamma_k&0&0&0&0 &\gamma_k \\
0 & -\gamma_k & - g_k & g_k & \gamma_k & -4J_3 \end{array} \right);~~
\end{eqnarray}
see Appendix \ref{tb} for further details.
\subsection{Numerical scheme}\label{num-scheme}
Numerical evolution of the time-dependent quantum states needs a careful algorithm to avoid the large error accumulation, which might result in producing a non-physical outcome. We use a numerical algorithm that captures the time evolution starting from an arbitrarily large time in the past to an arbitrarily large time in the future and proceeds in discrete but small steps $dt$. The unitary operator describing the time evolution is given by,
\begin{equation}\label{evolv}
\hat{U} (k, t) = \left( \hat{I} + i \hat{h}(k, t) \frac{dt}{2}\right) \left( \hat{I} - i \hat{h}(k, t) \frac{dt}{2}\right)^{-1},
\end{equation}
where $\hat{h}(k, t)$ is the Hamiltonian matrix governing the time evolution which is (\ref{4state-coupled}) in the four-level setup and is (\ref{6state-coupled}) in the six-level one, with the quench protocol mentioned earlier. The identity matrix $\hat{I}$ is a $4\times 4$ or a $6 \times 6$ one depending on the set-up. Such a time evolution operator has been used to study a four-level L-Z model earlier in the context of the model Hamiltonian \cite{Sin1, Sin2}. The evolution operator \eqref{evolv} is equivalent to a true evolution operator in the limit of $dt \rightarrow 0$, and reduces the error to $\frac{(dt)^2}{4}$ (reduction by a factor of 4) at $\mathcal{O}(dt^2)$. In this paper, we calculate the following quantities, $\langle 0| \hat{U} (k, t) |0\rangle $ and $\langle f| \hat{U} (k, t) |0\rangle $, where $|f\rangle$ are the final states except the post-quench ground state (i.e., the ground state corresponding to $t=\infty$).
The quantity $P_{00} = \langle 0| \hat{U} (k, t) |0\rangle$ describes the probability of the system to remain in the initial state. Using \eqref{4state-coupled} (for the four-level problem), \eqref{6state-coupled} (for the six-level problem), and \eqref{evolv} we determine the defect density $ n_d = \frac{1}{\Omega_{HBZ}} \int_{HBZ} d^3 k ~ P_{00} $, with $ \Omega_{HBZ} = \pi^3 / 2 $ being the volume of the HBZ.
\paragraph*{}
Furthermore, we aim to determine the dependence of the defect density on the quench rate $1/\tau$ (or equivalently $1/J\tau$). Physically, the defect density refers to the number density of the created excitations \cite{Dam, Pol2, Pol3}. These excitations are generated once the system is taken across a QCP where the system fails to remain in the adiabatic regime. As explained above, depending on the choice of the initial state the 3D-Kitaev model reduces to a four-level or a six-level L-Z problem for each momentum $\mathbf{k}$.

\section{Analytical scheme- the independent crossing approximation}\label{ica}

 We use the $ independent~ crossing~ approximation$ ($\rm{ICA}$)~\cite{Sin1} to analytically solve the multi-level L-Z problem. This treats the multi-level L-Z problem as a collection of independent two-level problems, each two-level problem being solved without considering the influence of the rest of the levels. We use the $\rm{ICA}$ to calculate the scattering matrix $S$ whose elements' squared $P_{ij} = |S_{ij}|^2$ represent the transition probabilities from $i$'th state at $-\infty$ to $j$'th state at $+\infty$.
\paragraph*{}
 Within the $\rm{ICA}$, one follows the diabatic levels, as illustrated in Fig.\ref{dia-adia-label}(a) and (c), and apply the two-level L-Z formula at each crossing point along the path \cite{Sin1, Sin2, Sin3, Landau, Zen}. To apply the $\rm{ICA}$, the avoided level crossing points must be far from each other, and this is indeed satisfied in our case as shown in Fig.~\ref{dia-adia-label}(b) and (d). During the quenching process when the system encounters the avoided level crossings it exhibits maximum probability of making transitions to the higher excited states. The necessary criteria for the applicability  of the $\rm{ICA}$ is the cancellation of the dynamic phase (gained along the path of evolution) in the expression of the transition probabilities \cite{Sin1, Sin2, Sin3}. It indeed turns out that for the four-level problem the dynamic phases gained along all the possible trajectories corresponding to the transitions from the initial state $|0\rangle$ get canceled in the expression of the probability density, as shown in Appendix \ref{ta}. However, in the case of trajectories forward in time, all the transitions from states $|1\rangle$ and $|2\rangle$ exhibit the effects of the dynamical phases.
 On the other hand, for the six-level problem the dynamic phase difference remains finite for a couple of paths corresponding to the transition from the state $|\tilde{0}\rangle$ (the ground state at $-\infty$), as shown in the Appendix \ref{ta1}.
\paragraph*{}
It is worthwhile to point out that the above approximation is semi-classical because this does not allow an evolution backward in time. Therefore, transitions from $|0\rangle$ to state $|1\rangle$, and from $|3\rangle$ to state $|2\rangle$ are not possible in the case of four-level problem.  However, for the six-level problem, transitions from state $|\tilde{0}\rangle$ to all other states are possible without requiring any evolution backward in time. In this sense, the six-level problem does not distinguish between the semi-classical and the quantum nature of the system. Further details of the calculations of the transition amplitudes, evaluated within $\rm{ICA}$, corresponding to both the four-level and six-level problems are described in Appendices \ref{ta} and \ref{ta1}, respectively.

\begin{figure}
\includegraphics[scale=0.175]{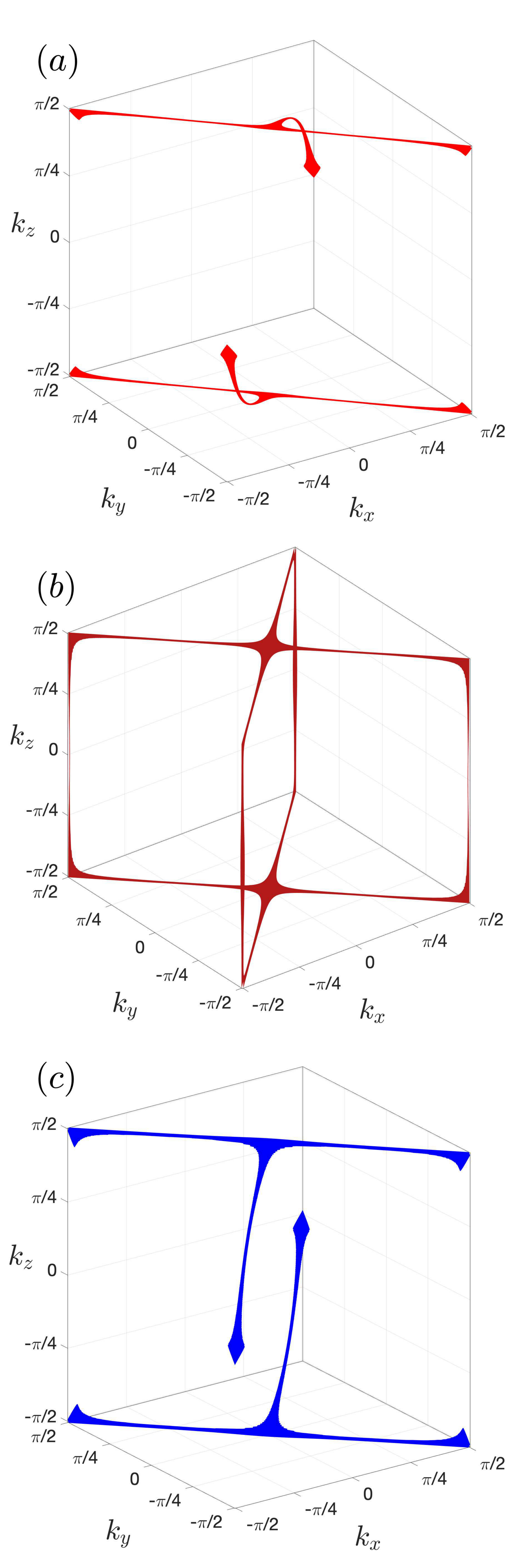}
\caption{\textbf{Non-adiabatic conditions for different $\alpha$:} Plot of the contour generated by the condition $|\Delta_{1k}|^2 = 0$. (a) for $\alpha=\pi /8$, (b) for $\alpha= \pi /4$, and (c) for $\alpha= 3\pi /8$. The set point $(k_{x}^{0}, k_{y}^{0}, k_{z}^{0})$, around which the Taylor expansion is made in \eqref{taylor}, lie on the contour. We further point out that from the numerical perspective implementation of the condition $|\Delta_{1k}|^2 = 0$ is done by considering a threshold which we take $10^{-6}$. This leads to the thickening of the lines corresponding to $|\Delta_{1k}|^2 = 0$.}
\label{contour_plot}
\end{figure}

To our surprise, we observe that the analytical expressions corresponding to the probability $P^{(i)}_{00}$ (with $i=4,6$ corresponding to the four and six-level problems obtained in Appendices \ref{ta} and \ref{ta1}, respectively) for the system to remain in the ground/initial state (of $t =-\infty$), within the semi-classical approximation, is very similar for both the four-level and the six-level problems. If we denote $p^{(4)}_k=P^{(4)}_{00}$ and $p^{(6)}_k=P^{(6)}_{00}$ as the transition probabilities within the ICA, we find that,
\begin{eqnarray}\label{P00}
p_{k}^{(6)} && = \exp \left( -\frac{\pi (J\tau)}{J^2} \label{4levpk} |\Delta_{1\mathbf{k}}|^2 \right),~~ p_k^{(4)}= \sqrt{p_{k}^{(6)}}, 
 \end{eqnarray}
where $p^{(4)}_k$ and $p^{(6)}_k$ differ by a factor of $1/2$ in the exponential. The defect density can be obtained straightforwardly  by using the following formula,
\begin{equation}\label{def}
n^{(i)}_d = \frac{1}{\Omega_{HBZ}} \int_{HBZ} d^3 k \; p^{(i)}_k,~~i=4,6,    
\end{equation}
where 
the integration is performed over the HBZ as indicated above. It is worthwhile to point out that the probabilities $P_{0j}^{(i)}$ for the system to reach a state $|j\rangle$ at $t=\infty$ (except the corresponding ground state) also represent the defect densities. However, we have checked numerically for both the four-level and the six-level problems that the defect densities corresponding to $P_{0j}^{(i)}$ exhibit the same scaling with the quench rate $1/\tau$ as that of \eqref{def}, thereby providing no new physics. Therefore, in what follows we evaluate \eqref{def} both analytically and numerically.
From \eqref{4levpk} and \eqref{def}, we  observe that the determining factor for the overall behavior of defect density $n^{(i)}_{d}$ is  $|\Delta_{1\mathbf{k}}|^2$, which has the following expression,
\begin{eqnarray}\label{Del}
|\Delta_{1\mathbf{k}}|^2 &&= 4J^2 \Bigg[ \cos^{2}\alpha \cos^{2} \frac{k_+}{2}+ \sin^{2} \alpha \, \cos^{2} \frac{k_-}{2} \nonumber  \\ && +  \sin 2\alpha \cos (k_x - k_y)  \cos \frac{k_+}{2} \cos\frac{k_-}{2}\Bigg],
\end{eqnarray}
where $k_{\pm}=k_x + k_y \pm 2 k_z$. For sufficiently slow quench $J\tau >> 1$, the quantity $p_k$ is exponentially small for all values of $\mathbf{k}$ except on a contour dictated by $ |\Delta_{1\mathbf{k}}|^2 = 0$ (see Fig. \ref{contour_plot}).
We henceforth refer to $ |\Delta_{1\mathbf{k}}|^2 = 0$, for our case, as the non-adiabatic condition \cite{KS1, KS2, Pol2, Pol3}, and the corresponding points (satisfying the above condition) in the HBZ as the non-adiabatic points. Physically, it means that for all the non-adiabatic points and the points near these, the system exhibits the maximum non-adiabatic nature, thereby maximizing the probability of formation of the defects.

Next we calculate the defect density using \eqref{P00} and \eqref{def} within the stationary phase approximation. We note that if a generic point on the above mentioned contour is  given by $(k_{x}^{0}, k_{y}^{0}, k_{z}^{0})$, the most dominant contributions to the momentum integrals come from the values of $\mathbf{k}$ close to these points.
Also, without loss of generality if we substitute $k_x = k_y \pm 2 n \pi$ (corresponding to $\cos(k_x - k_y) = 1$ in \eqref{Del} with $n$ being integers), we obtain from  $|\Delta_{1\mathbf{k}}|^2 = 0$,  
\begin{eqnarray}\label{min_cond}
\tan(k_x) \tan(k_z) &=& \frac{1+\tan \alpha}{1-\tan \alpha}, k_x = k_y \pm 2 n \pi 
\end{eqnarray}
where $\alpha=J_x/J_y$. The above equation indeed represents a set of critical lines (or, $d-m =1$ dimensional critical lines) for each values of $\alpha$ indicating $m=2$ in our 3D ($d=3$) model. However, the above conditions are valid only for all the values of $\alpha$ except $\alpha = 0, \; \text{and}\, \pi/2$. For these two values of $\alpha$ the non-adiabatic condition is needed to be obtained directly from \eqref{Del}. It is therefore possible to Taylor-expand $|\Delta_{1\mathbf{k}}|^2$ around $(k_{x}^{0}, k_{y}^{0}, k_{z}^{0})$ on the above mentioned set of lines.

By making the above mentioned expansion up to leading order (i.e., the stationary phase approximation), we obtain 
\begin{eqnarray}\label{taylor}
|\Delta_{1\mathbf{k}}|^2 &=& \frac{1}{2}(4J^2)\Big[4f_{xx} (k_x - k_{x}^{0})^2 + 4f_{xz}(k_x - k_{x}^{0})(k_z - k_{z}^{0}) \nonumber \\ &+& f_{zz}(k_z - k_{z}^{0})^2 \Big],
\end{eqnarray}
where we have rewritten $\frac{\partial^2 (|\Delta_{1\mathbf{k}}|^2)}{\partial k_i \partial k_j} = 4J^2 f_{ij} (\alpha)$. The quantity $\left[ \frac{\partial^2 (|\Delta_{1\mathbf{k}}|^2)}{\partial k_i \partial k_j} \right]$ is the $ij$'th element of the precision matrix $(2\times 2)$ of a multivariate Gaussian. For $J\tau >>1$, the values of $p_k$ corresponding to \eqref{P00} is turn out to be very small for all values of $\mathbf{k}$ except the points near $(k_{x}^{0}, k_{y}^{0}, k_{z}^{0})$. We can then extend the upper limit of $k$-integration to $\infty$, and apply Gaussian integral in \eqref{def}. In deriving the \eqref{taylor}, we have used the condition $k_x = k_y \pm 2n \pi$, i.e., $k_x$ and $k_y$ are related to each other by a constant shift. This translates into  $k_x - k_{x}^{0} = k_y - k_{y}^{0}$ which reduces the precision matrix to a $2\times 2$  from a $3 \times 3$ one. 
 Evaluating the Gaussian integral we find the defect density as, 
\begin{equation}
n_d = \frac{\pi}{\Omega_{HBZ}} \left( \frac{1}{(J\tau) \sqrt{|| f_{ij}(\alpha) ||}}\right),
\end{equation}
where $|| f_{ij}(\alpha) || = 4[f_{xx} f_{zz} - (f_{xz})^2]$ is the determinant of the $2\times2$ precision matrix, which must be positive definite. It turns out that for $\alpha = 0\; \text{and}\; \pi/2$ the above mentioned determinant vanishes and we can't evaluate the Gaussian integral. Therefore, we have evaluated the corresponding integral in \eqref{def} numerically within the HBZ. Our analytical calculation using $\rm{ICA}$ thus establishes $ n_d \propto (J\tau)^{-1}$, in the limit of the very slow quenching.

\section{Defect density}\label{defect}
\begin{figure}
\includegraphics[scale=0.53]{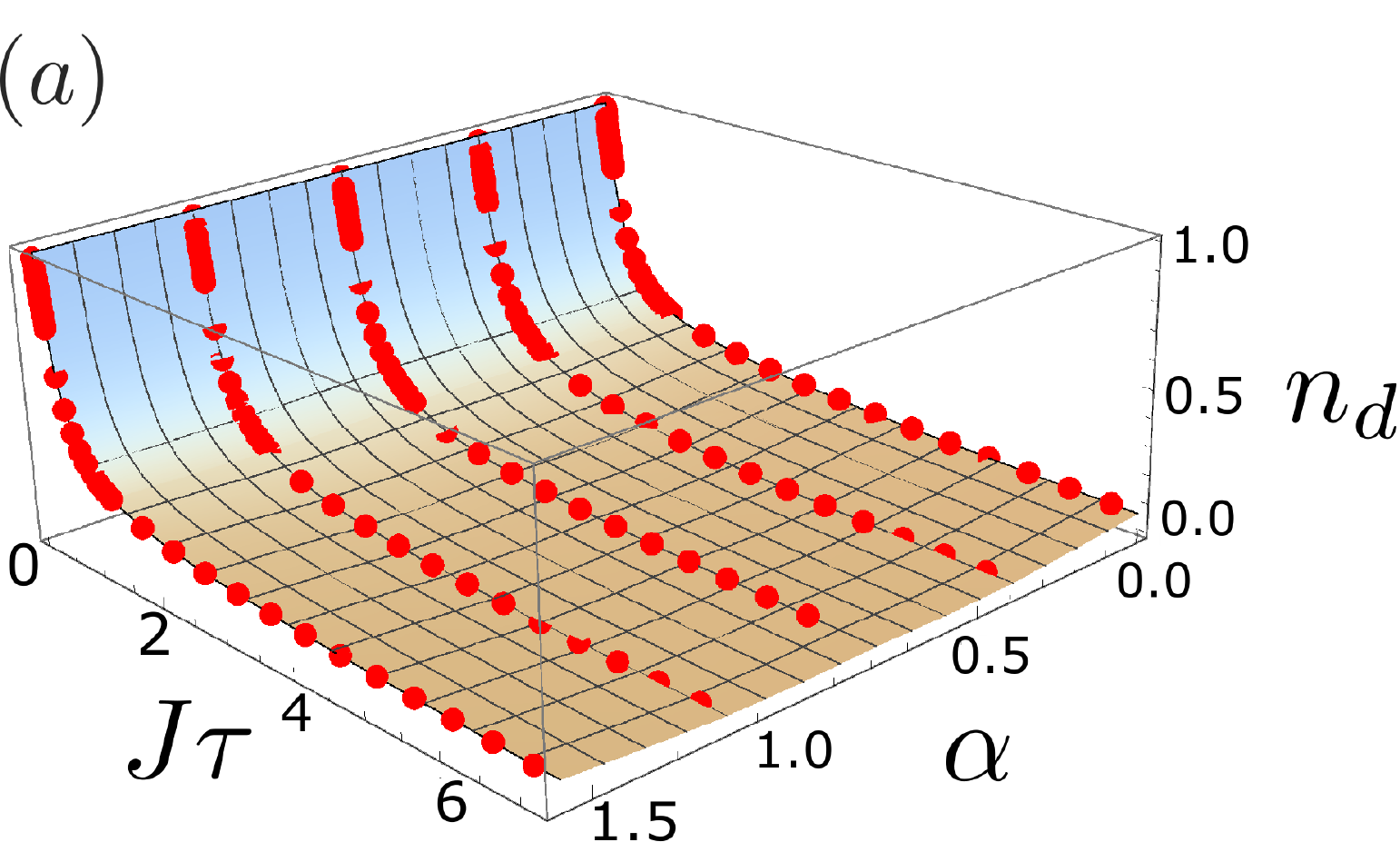} \\
\includegraphics[scale=0.36]{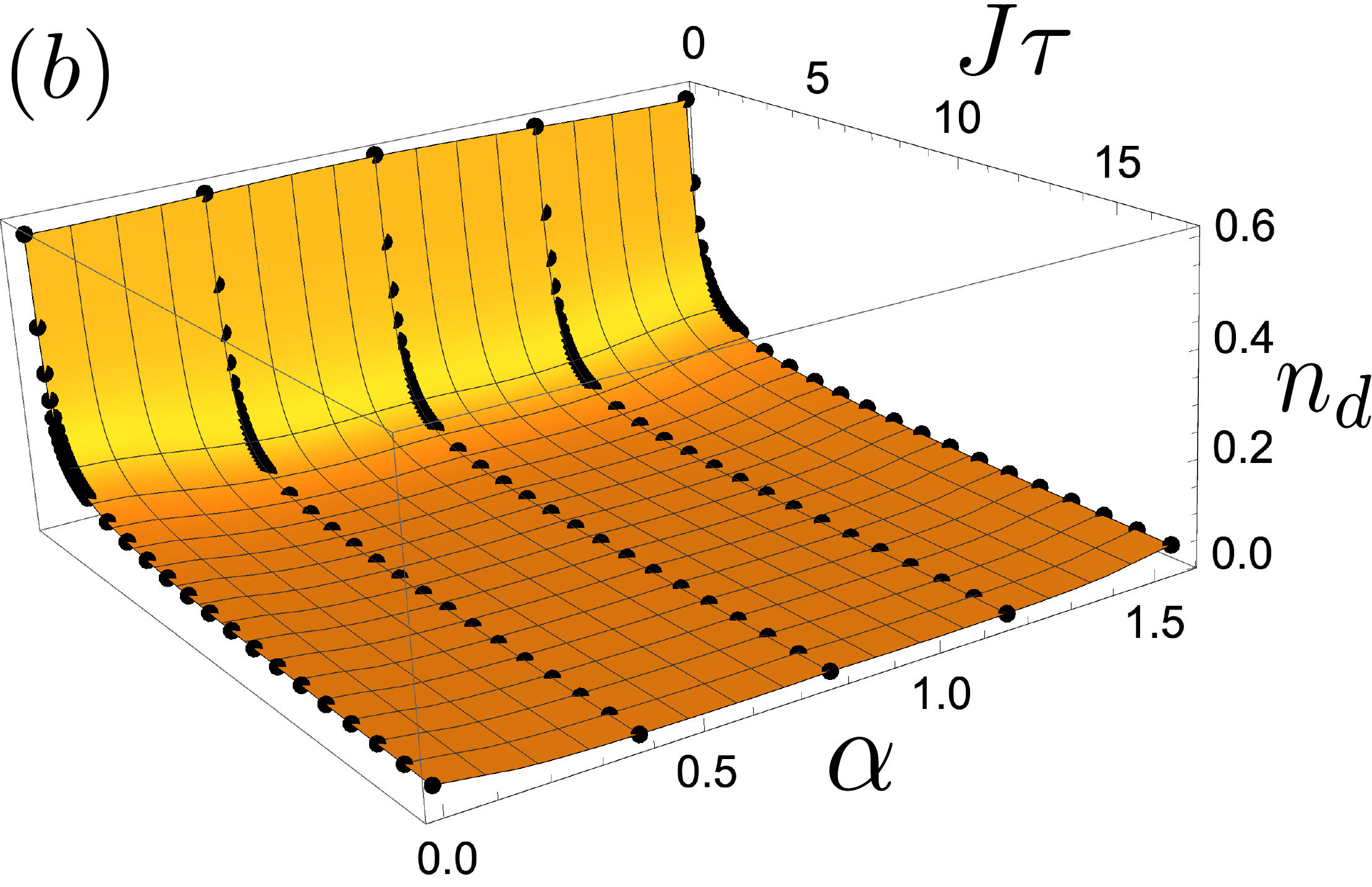}
\caption{\textbf{Defect density:} Plot of defect density $n_d$ as a function of $J\tau$ and $\alpha$ where $\alpha = \tan^{-1} (J_y / J_x)$. (a) For the four-level problem, the red dotted lines are the results corresponding to the exact numerical calculations, and the surface plot is the result corresponding to the ICA. (b) For the six-level problem, the black dotted lines are the results corresponding to the exact numerical calculations, and the surface plot is the result corresponding to the ICA.}
\label{defect-new}
\end{figure}

In Fig. \ref{defect-new}(a) and \ref{defect-new}(b), we have plotted the defect density with respect to $J \tau$ and $\alpha$ for the four-level and the six-level L-Z problems respectively. The dots in Fig.~\ref{defect-new} are the numerically (following the scheme outlined in \ref{num-scheme}) obtained defect density. The 3D plot is obtained by using \eqref{def}.  We find that both the plots agree with each other with high degree of accuracy implying the validity of $\rm{ICA}$ in this model for both the four-level and the six-level problem. For slow quenching in both cases we obtain $n_d \sim \tau^{-1}$. The Appendix \ref{ta3} features the `log-log' plot (see Fig. \ref{fit} (a)) for $n_d$ vs. $J\tau$, which shows that for $\alpha = \pi/8, ~ \pi/4$ and $3\pi/8$ the scaling is indeed $\tau^{-1}$. It is worthwhile to point out that for these values of $\alpha$'s the system exhibits a 3D structure. For $\alpha = 0,~ \text{and} ~ \pi/2$ the system becomes a set of independent Kitaev chains, and Fig. \ref{fit} (b) shows that $n_d$ scales as $1/\sqrt{\tau}$ as expected.
\paragraph*{}
Next we compare the above results with the prediction of Ref. \cite{KS1} that defect density crucially depends on the `dimensionality' of the critical hyper-surface through  which the system passes during the quenching. The important quantities are $z$, which tells us how the energy dispersion of a given system depends on momentum at low energy, the dynamical exponent $\nu$, which determines the asymptotic dependence of two point correlation function at large distance, and $m$, which tells us how the system becomes gapless in momentum space. In the case of our 3D Kitaev model, it can be checked easily that $\nu = z = 1$. The dispersion of the 3D Kitaev model is such that it vanishes on a contour which constitutes a 2D critical hyper-surface under the action of quench yielding $m=2$. Keeping in mind that our system is a 3D one with $d=3$, and substituting the above values of $(z,\nu,d,m)$ in the expression of defect density $n_d \propto \tau ^{-\frac{m \nu}{z \nu + 1}}$, we obtain $n_d \propto \tau^{-1}$. It is worthwhile to point out that the 2D Kitaev model exhibits a scaling $n_d \propto 1/\sqrt{\tau}$ \cite{KS1}.
\paragraph*{}
We now discuss the gapless condition of the spectrum and its relation to the scaling of defect density. Diagonalization of \eqref{4LZ} yields the following eigenvalues,
\begin{eqnarray}
\mathcal{E}^{1-4}_k= \pm \sqrt{8 J_z^2 + 2 (|\Delta_{1k}|^2 +  |\Delta_{2k}|^2 ) \pm \Delta_k  },
\end{eqnarray}
where $\Delta_k= 4|\Delta_{2k}| \sqrt{4 J_z^2 + |\Delta_{1k}|^2 \cos^2 \theta_{2k} }$. The gapless condition of the spectrum is found to be,
\begin{eqnarray}
&& \left( 4 J^2_z + |\Delta_{1k}|^2 - |\Delta_{2k}|^2  \right)^2  \nonumber \\
&&+ 4 |\Delta_{1k}|^2 |\Delta_{2k}|^2 \sin^2 \theta_{2k}=0.~~~\label{fourgapless}
\end{eqnarray}
We  note from the above equations that the condition $|\Delta_{1k}|^2=0$, i.e., the non-adiabatic condition is a subset of the gapless condition. Thus our analysis suggests that for a general multilevel problem with partial or complete filling of the negative energy states the scaling of the defect density is decided by the coupling of the relevant levels in a given system.

\begin{figure}[h]
\includegraphics[scale=0.33]{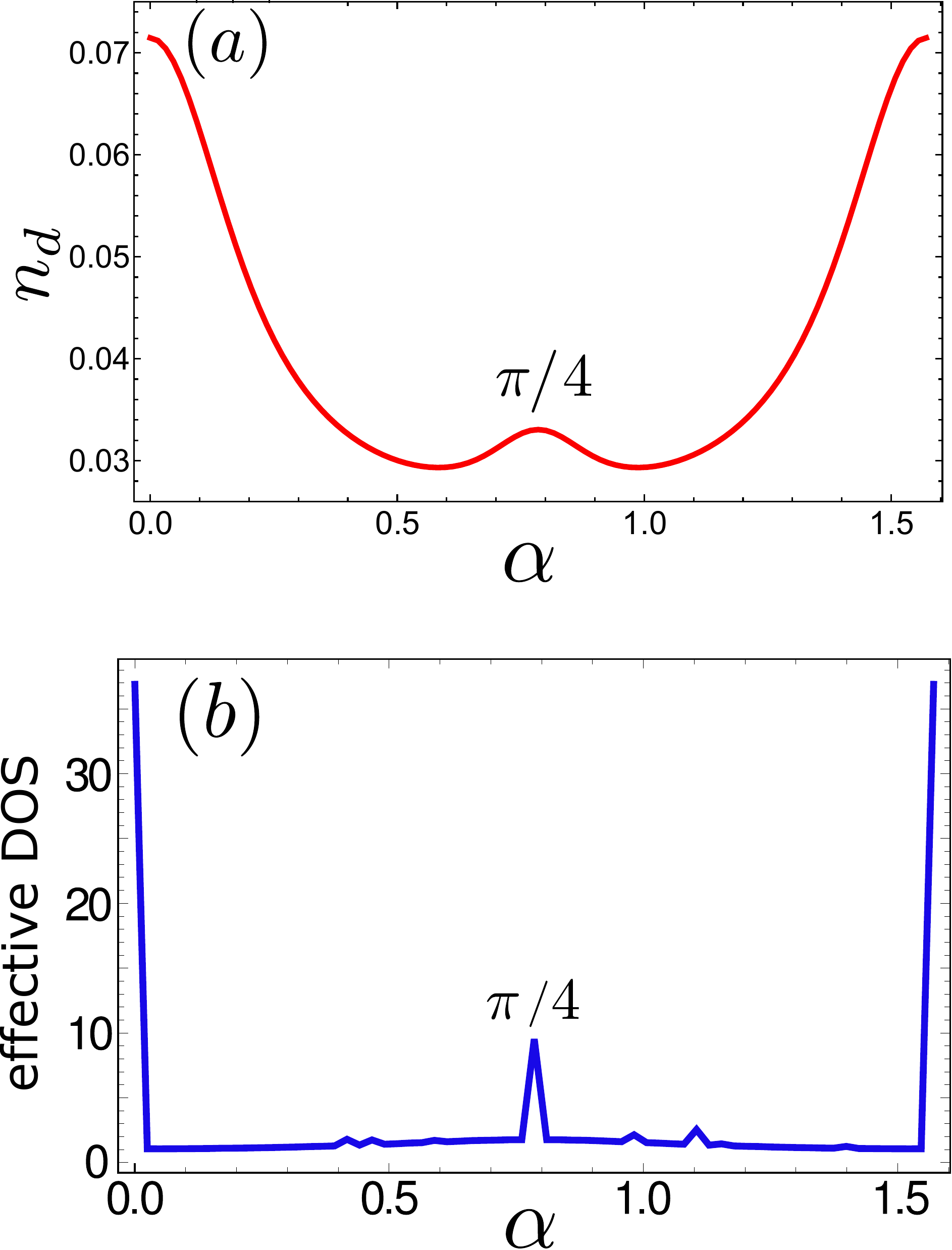}
 \caption{\label{defect_alpha} \textbf{ Effective-DOS plot:} (a) Plot of  defect density $n_d$ as a function of $\alpha$ where $\alpha = \arctan (J_y / J_x)$ for $J\tau >> 1$. The red dotted points are results corresponding to exact numerical calculations, and the surface plot is the result corresponding to the Independent crossing approximation. A hump at $\alpha = \pi/4$ signifies a local maximum. (b) Effective Density of States (DOS) as a function of $\alpha = \tan^{-1} \frac{J_y}{J_x} $ at the gapless phase.}
\label{effective_dos}
\end{figure}

However, the reason that the scaling of the defect density follows the law proposed in Ref. \cite{Pol2} is that for the case of the 3D Kitaev model the conditions for the zero of the spectrum, and the condition $|\Delta_{1k}|^2 = 0$ both vanish on a contour in an identical manner, (i.e., having $m=2$). We note that this is purely coincidental for the case of 3D Kitaev model.
\paragraph*{}
Remarkably, the asymptotic behavior of defect density (i.e., the behavior of $n_{d}$ in the limit $J\tau >> 1$) corresponding to the 2D Kitaev model differs from that of the 3D Kitaev model with respect to it's dependence on $\alpha$. In the case of the 2D Kitaev model, when  $\alpha$ is varied from 0 to $\pi/2$, the defect density increases monotonically up to $\alpha=\pi/4$ and then decreases monotonically such that it is symmetric with respect to $\alpha=\pi/4$ \cite{KS1}. For the 3D Kitaev model although the asymptotic behavior of the defect density for large $J\tau$ is symmetric with respect to $\alpha=\pi/4$ however, in contrast to the 2D Kitaev model, it exhibits the maxima at $\alpha=0$ and $\pi/2$, and decreases monotonically and subsequently increases to reach local maximum at $\alpha=\pi/4$ as shown in Fig. \ref{defect_alpha} (a).
  
 To explain this unusual feature we look back to the expression of $p_k$ appearing in \eqref{P00} and \eqref{def}.
 Therefore, it is reasonable to define an effective density of states (effective-DOS) corresponding to the variable $\Delta_{1k}$, which essentially quantifies the number of $\mathbf{k}$-points in HBZ satisfying the non-adiabatic condition $\Delta_{1k}=0$. 
 
 From the effective-DOS plot in Fig.~\ref{defect_alpha}(b) we see that the effective DOS corresponding to $\Delta_{1k}=0$ is maximum at $\alpha=0~ \text{and}~ \pi/2$, and decreases initially away from these two values of the $\alpha$. However, the effective DOS exhibits a local maximum at $\alpha=\pi/4$. The infinite effective-DOS at $\alpha =0$ and $\pi / 2$ refer to the contribution from each of the decoupled chains which are infinite in number. This explains the variation of asymptotic values of $n_d$ with respect to $\alpha$ in the sense that the more the number of $\mathbf{k}$-points in HBZ satisfy the non-adiabatic condition, the more the defect is generated.

\section{Correlation functions}\label{correl}
In this section, we discuss the effect of the quenching in the correlation function and its connection to defect production. We note  that the true quantum ground state of the 3D Kitaev model is supposed  to be a quantum spin liquid where the spin-spin correlation (two spin correlation) function is short-ranged and anisotropic \cite{SM-GB}. The spin-spin correlation is non-zero only for the nearest neighbor spins for all values of the parameters of the system. It remains an intriguing question how the correlation function develops non-locality due to the quenching. For simplicity we consider the following two-spin correlation function operator,
\begin{equation}
\mathcal{O}^{3D} (\mathbf{r}) = \sigma_{\mathbf{R}}^{z} \sigma_{\mathbf{R+r}}^{z},
\end{equation}
which is nothing but a product of the spin operators corresponding to a site $\mathbf{R}$ at $a$ sub-lattice and a site $\mathbf{R+r}$ at $b$ sub-lattice. The quantity $\mathcal{O}^{3D} (\mathbf{r})$ is non-zero only for $\mathbf{r}=0$, and this fact remains the same for all points in the parameter space.  In the Majorana fermion representation, the above correlation function takes the form $\mathcal{O}^{3D} (\mathbf{r}) =   S_{\mathbf{R},\mathbf{R+r}}i c_{1a}(\mathbf{R}) c_{1b}(\mathbf{R+r}) $ where $S_{\mathbf{R},\mathbf{R+r}}$ is the product of the $\mathbb{Z}_2$ gauge fields defined on the bonds along the path. In the following, we study the evolution of this two-fermion correlation function operator $ i c_{1a}(\mathbf{R}) c_{1b}(\mathbf{R+r}) $ rather than the spin-spin correlation. Thus the object of our interest is, 
\begin{equation}
\label{cacb}
\mathcal{C}^{3D} (\mathbf{r}) = i c_{1a}(\mathbf{R}) c_{1b}(\mathbf{R+r}).
\end{equation}
Doing a Fourier transform, the ground state expectation value of the two fermion correlation takes the following form,
\begin{equation}
\label{cacbk}
\langle \mathcal{C}^{3D} (\mathbf{r}) \rangle= \pm \frac{1}{2N} \sum_{k \in HBZ} \cos(\mathbf{k \cdot r}),
\end{equation}
where $+$ and $-$ signs refer to the ground state (or initial state in the case of the four-level problem) corresponding to $J_z= -\infty$ and $\infty$ respectively and $\langle \mathcal{C}^{3D} (\mathbf{r}) \rangle$ is indeed $\pm \delta_{\mathbf{r}, 0}$. The derivation of Eq. \ref{cacbk} is similar to previous quench study in 1D and 2D Kitaev models \cite{KS2}, and has been briefly described in Appendix \ref{tbc}.

The final state after quench is   superposition of  both the initial ground state and excited states such that $p_k$ represents the probability of finding the system in the ground state of $t=-\infty$ (which is an excited state corresponding to $t=\infty$). In the limit $J_z \rightarrow -\infty$, the system is in the ground state where nearest-neighbor spins are ferromagnetically aligned and, as the quenching takes place, the state of the system deviates from the ground state configuration (corresponding to $t=-\infty$) and the defects are generated. Physically, $p_k$ quantifies how much ferromagnetic component is present in the final  state given the fact that the initial state at $t=-\infty$ was a ferromagnetic state.
In this case, the correlation function takes the form,
\begin{equation} \label{corr_r}
\langle \mathcal{C}^{3D} (\mathbf{r}) \rangle=  -\delta_{\mathbf{r}, 0} + \frac{1}{\Omega_{HBZ}} \int d^3 k p_k \cos(\mathbf{k \cdot r}),
\end{equation}
where the second term in the above equation is a measure of defect correlation (i.e., the two-spin correlation in the presence of defects) and henceforth will be analyzed. 
We note that the final state characterizes the excitations or defects generated in the system due to the quench and, therefore, we refer the correlation function in the quenched state as the defect correlation following the earlier usages \cite{KS2}.

We evaluate the second term of the above equation by expanding the defect probability $p_k$, obtained within the $\rm{ICA}$, in the limit of very slow quench corresponding to $J\tau >> 1$, for both four and six-level problems. As usual. the dominant contribution comes from the contour determined by $|\Delta_{1\mathbf{k}}|^2 =0$. Evaluating the relevant Gaussian integrals, as has been done for the defect density in the previous section, we find,
\begin{eqnarray}\label{correl_r}
& & \langle\mathcal{C}^{3D}(\mathbf{r}) \rangle = \frac{\pi}{\Omega_{HBZ}}\left( \frac{\cos\left[k_{x}^{0} (N_{x}+N_{y}) + k_{z}^{0} N_z\right]}{(J\tau) \sqrt{|| f_{ij}(\alpha) ||}}\right) \times \nonumber \\  & &\exp \left[ -\frac{(N_{x}+N_{y})^2 f_{zz}+ 4N_{z}^{2} f_{xx} - 4(N_{x}+N_{y}) N_{z} f_{xz}}{4\pi J\tau || f_{ij}(\alpha) ||} \right] . \nonumber \\ 
\end{eqnarray}
The above equation shows that the asymptotic behavior of the defect correlation as a function of $J\tau$ is given by,
\begin{equation}
\label{cortau}
\langle\mathcal{C}^{3D}(\mathbf{r}) \rangle \sim \tau ^{-1} \exp(-A \tau^{-1}),
\end{equation}
where $A = \frac{(N_{x}+N_{y})^2 f_{zz}+ 4N_{z}^{2} f_{xx} - 4(N_{x}+N_{y}) N_{z} f_{xz}}{4\pi J || f_{ij}(\alpha) ||}$, and $N_x = 2n_1  + n_3$, $N_y = 2n_2 + n_3$, and $N_z = 2n_3$. The above asymptotic behavior, therefore, remains valid for both the four and the six-level problem.

\paragraph*{}
\subsection{Comparison of Defect correlation in various planes}
After obtaining the asymptotic analytic expression of defect correlation function in \eqref{correl_r}, we analyze the latter in different directions and planes in the real-space. We evaluate (\ref{corr_r}) numerically in various planes and specific directions by noting that any arbitrary point on the lattice is given by $\mathbf{r} = n_1 \mathbf{a}_1 + n_2 \mathbf{a}_2 + n_3 \mathbf{a}_3$. First we begin by discussing how the spatial correlation builds in certain planes.
\begin{figure}[h]
\includegraphics[scale=0.34]{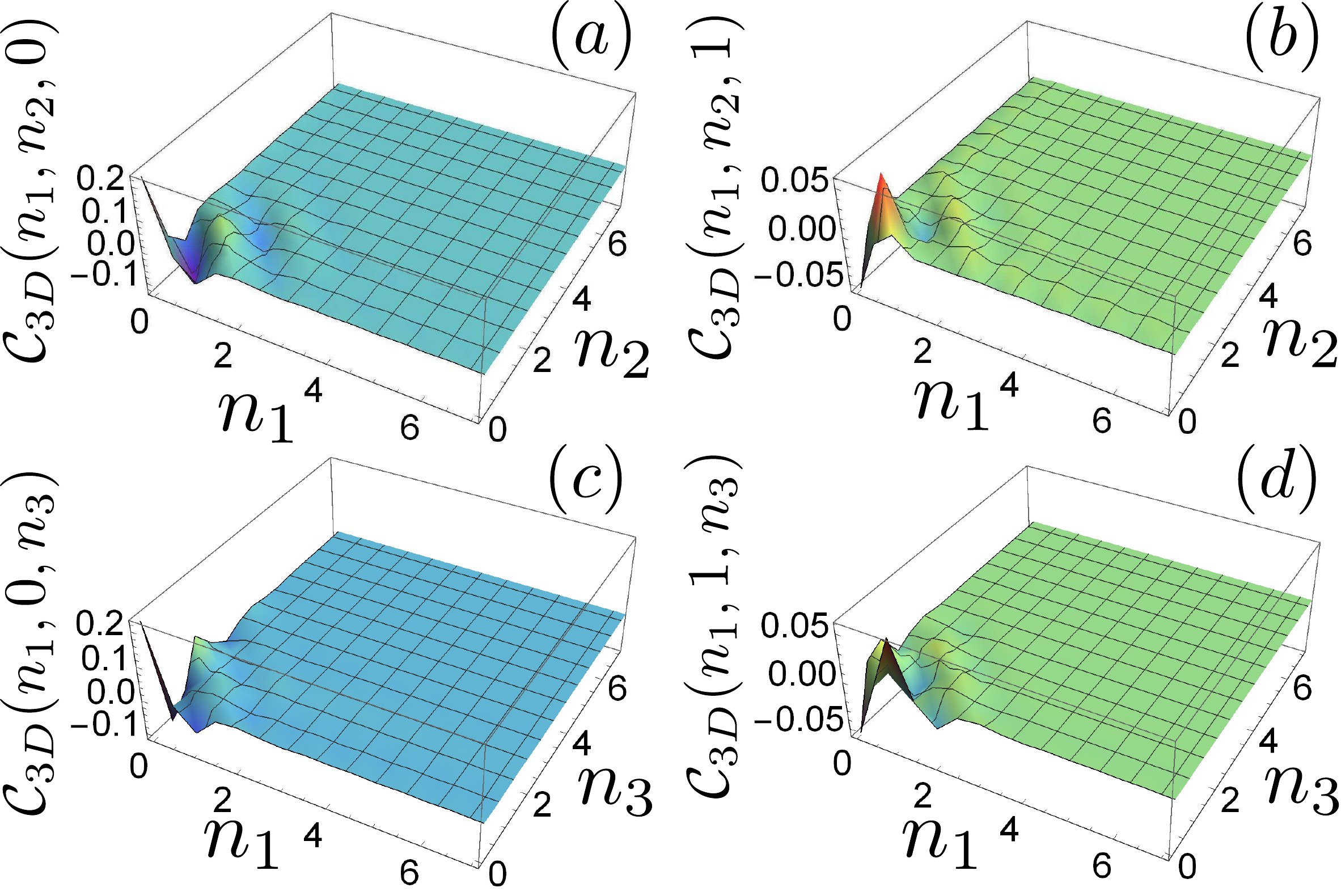}
\caption{\label{corr-fig-six}\textbf{Spatial variation of the correlation for the four-level problem:} Plot of defect correlation $\langle\mathcal{C}^{3D}(\mathbf{r}) \rangle$ corresponding to the four-level representation as a function $n_1, n_2$ and $n_3$, i.e., as a function of spatial coordinate $\mathbf{r}$. The spatial variation of the defect correlation in the $a_1 - a_2$-plane (a) when $n_3 = 0$ and (b) when $n_3 = 1$. The spatial variation of the defect correlation in the $a_1 - a_3$-plane (c) when $n_2 = 0$ and (d) when $n_2 = 1$. See text for explanations.}
\end{figure}
\begin{figure}
\includegraphics[scale=0.36]{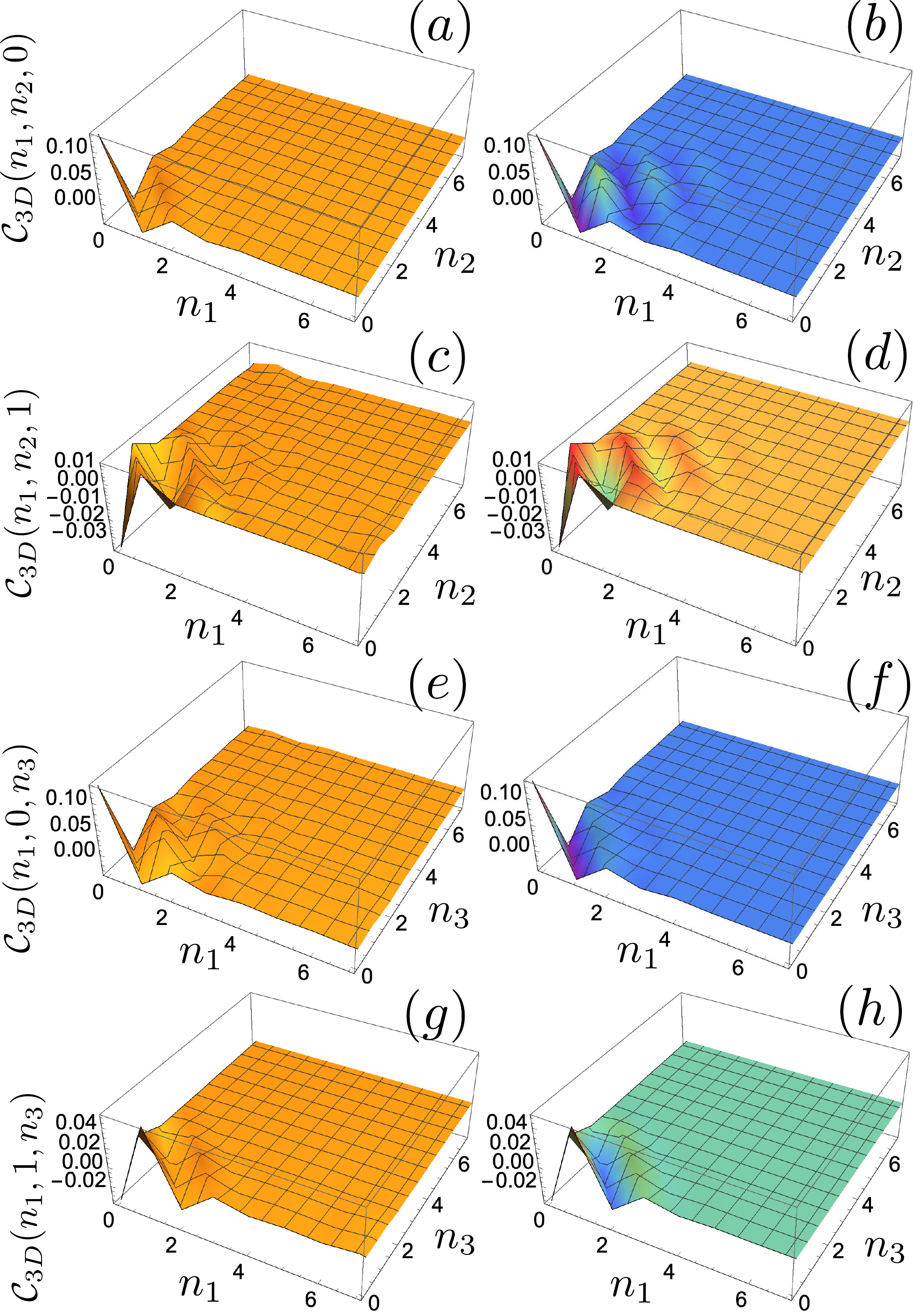}
\caption{\label{corr-fig}\textbf{Spatial variation of the correlation for the six-level problem:} Plot of defect correlation $\langle\mathcal{C}^{3D}(\mathbf{r}) \rangle$, corresponding to the six-level representation, as a function $n_1, n_2$ and $n_3$, i.e., as a function of spatial coordinate $\mathbf{r}$. (a), (c), (e), (g) are obtained from the full numerical calculation; (b), (d), (f), (h) are obtained from the ICA by using $p_k$ of (\ref{corr_r}) corresponding to the ICA. The spatial variation of the defect correlation in the $a_1 - a_2$-plane are plotted in (a) \& (b) when $n_3 = 0$, and in (c) \& (d) when $n_3 = 1$. The spatial variation of the defect correlation in the $a_1 - a_3$-plane are plotted in (e) \& (f) when $n_2 = 0$, and in (g) \& (h) when $n_2 = 1$. See text for explanations.}
\end{figure}

In Figs.~\ref{corr-fig-six}, and \ref{corr-fig} we plot the spatial variation of the defect correlation as a function of $n_1$, $n_2$, and $n_3$ for $J_x =J_y = J =1$ (i.e., $\alpha = \pi/4$), and also $J\tau =1$ for both the four-level and the six-level problems respectively.  For the four-level problem $\langle\mathcal{C}^{3D}(\mathbf{r}) \rangle$ has been evaluated by exact numerical calculation and for the six-level problem the same has been evaluated by both  exact numerical as well using  ICA for comparison. In Fig. \ref{corr-fig}, in the left panel we present the results obtained by exact numerical evaluation and in the right panel we present the same obtained within  ICA. For a detailed description we refer to the caption of Fig. \ref{corr-fig}. (We have not used ICA for the four-level problem to avoid redundancy.)  

\begin{figure}[h]
\includegraphics[scale=0.17]{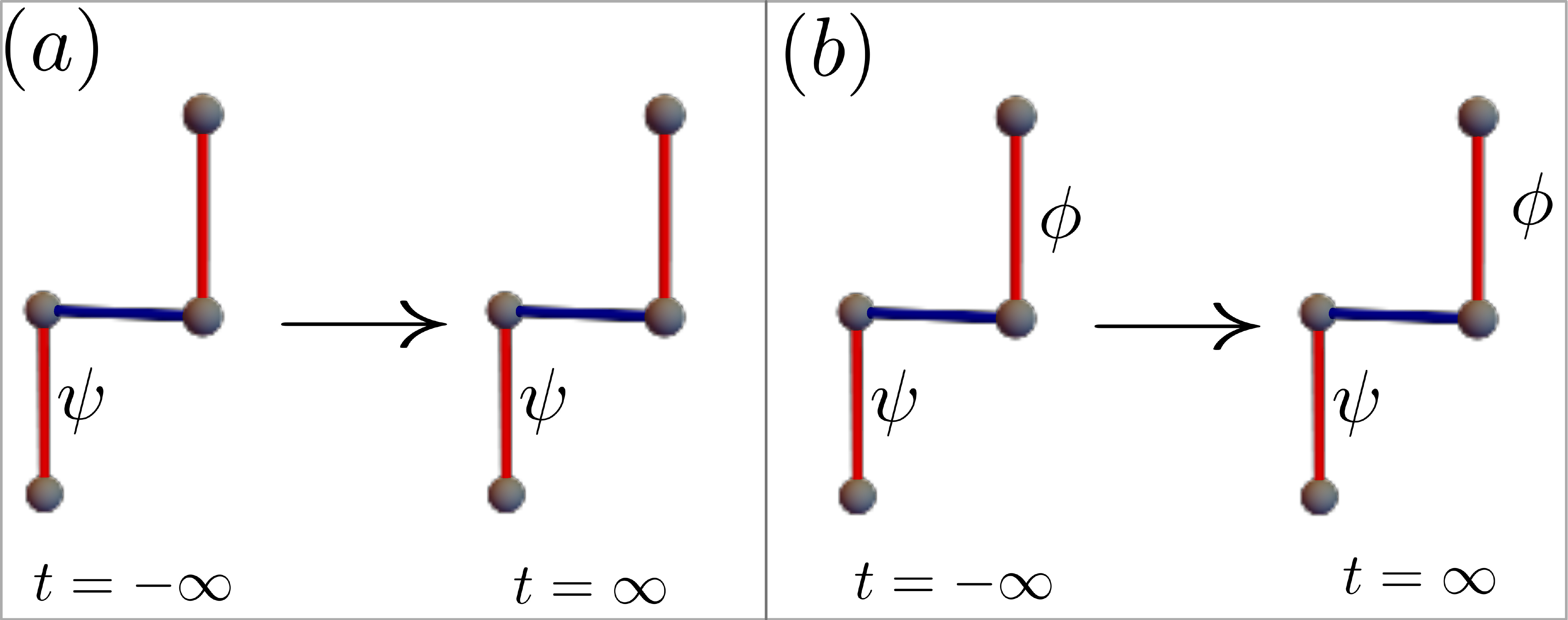}
\caption{\label{unitcell} \textbf{Unit-cell, and ground state configuration in terms of fermions:} Schematic picture of the ground state configuration considered (a) for the four-level problem with only the $\psi$ fermion state being occupied, and (b) for the six-level problem where both the $\psi$ and the $\phi$ fermions are occupied. These fermions are defined above \eqref{hkmat}.} 
\end{figure}

Physically, the difference between the four-level and  six-level problems concerning the correlation function study is that in the former the ground state of the system is such that only the $\psi$ fermion defined on the lower z-bond of the unit cell is occupied; see Fig. \ref{unitcell} for schematic illustration. On the other hand, for the six-level problem the ground state corresponds to both the $\psi$ and $\phi$ fermions defined on the both the z-bonds of the unit cell being occupied. We find that the correlation functions obtained in these two cases do not differ qualitatively because though in the four-level problem initially the $\phi$ fermion is absent, it affects the evolution of the $\psi$ fermion through the non-zero coupling $\Delta_{2k}$.

\subsubsection{Correlation function in the $\mathbf{a}_1-\mathbf{a}_2$ plane}
In Figs. \ref{corr-fig-six}(a) and \ref{corr-fig-six}(b), we plot defect correlation in the $\mathbf{a}_1-\mathbf{a}_2$ plane for $n_3=0$ and $n_3=1$ respectively for the four-level problem. Corresponding plot for the six-level problem are presented in Figs. \ref{corr-fig}(a) and \ref{corr-fig}(c) respectively.  These figures indicate that the nearest neighbor correlations are the dominant ones and the next nearest and other long range correlations with far away sites quickly die out. However, occasionally the correlation with the far away sites build up due to the lattice geometry and the connectivity. It is worth noticing that the spatial variation of the defect correlation is symmetric in the $\mathbf{a}_1 - \mathbf{a}_2$ plane with respect to the line $n_1=n_2$. 
Figures \ref{corr-fig}(b) and \ref{corr-fig}(d) represent the same correlation functions (plotted in  Figs. \ref{corr-fig}(a) and \ref{corr-fig}(c)) obtained using ICA. We notice that ICA is able to show more oscillations than the exact numerical evaluation as the latter method has some practical limitations.

\subsubsection{Correlation function in the $\mathbf{a}_1-\mathbf{a}_3$ plane}
In Figs.~\ref{corr-fig-six}(c) and \ref{corr-fig-six}(d), we plot the defect correlation in the $\mathbf{a}_1-\mathbf{a}_3$ plane for $n_2=0$ and $n_2=1$, respectively, for the four-level problem. The corresponding plots for the six-level problem are presented in Figs. \ref{corr-fig}(e) and \ref{corr-fig}(f), respectively. When comparing with the defect correlation in the $\mathbf{a}_1-\mathbf{a}_2$ plane, we find that in the $\mathbf{a}_1-\mathbf{a}_3$ plane the defect correlation is mostly zero and also not symmetric with respect to the line $n_1=n_3$ line.

\subsubsection{Correlation function in the $\mathbf{a}_2-\mathbf{a}_3$ plane}
Furthermore, owing to the symmetric nature of it in the $\mathbf{a}_1 - \mathbf{a}_2$ plane the spatial variation of the defect correlation in the  $\mathbf{a}_1 - \mathbf{a}_3$ plane is the same as that of the  $\mathbf{a}_2 - \mathbf{a}_3$ plane. It is worthwhile to point out that the symmetric nature of the correlation function arises due to the condition $\alpha=1$ (or $J_x=J_y$). Therefore, we expect the defect correlation to develop  spatial anisotropy for arbitrary values of the coupling constants corresponding to $J_x \neq J_y$.
\paragraph*{}

\subsection{Defect correlation along various directions}

To investigate the above mentioned anisotropic nature of the correlation function in real space when the ratio $J_y / J_x$ is varied from $0$ to $\infty$, while maintaining $J_{x}^{2}+J_{y}^{2} = 1$, we plot in Fig.~\ref{corr-alpha}(a) and Fig.~\ref{corr-alpha}(b), the correlation $\langle\mathcal{C}^{3D}(\mathbf{r}) \rangle$ as a function of $\alpha = \tan^{-1} \frac{J_y}{J_x}$ for the four-level and six-level L-Z quenches respectively. Here the quantity $\langle\mathcal{C}^{3D}(\mathbf{r}) \rangle$ has been obtained by numerically evaluating the second term of \eqref{corr_r} where $p_k$ has been obtained using both the $\rm{ICA}$ and the exact numerical calculations explained in \ref{num-scheme}.  All the correlation functions are the nearest-neighbor ones and obtained by substituting ${\bf{r}}=n_1 \mathbf{a_1}+ n_2 \mathbf{a_2} + n_3 \mathbf{a_3}$. Below we represent $\mathbf{r}$ by simply $(n_1,n_2,n_3)$ and discuss defect correlation for some representative combinations of  $(n_1,n_2,n_3)$. 
\subsubsection{Correlation function for $(1,0,0)$.}
First we discuss the behavior for ${\bf{r}}=(1,0,0)$ which describe how the correlation function behaves along $x$-direction. It is plotted in Fig. \ref{corr-alpha}(a) and (b) in brown for the four-level and six-level problems, respectively. This represents a nearest neighbor correlation function joined by the $z-y$ bonds along the $x$-direction. For $\alpha=0$ (i.e., $J_y=0$) these two bonds are disconnected and we obtain a zero correlation. In this limit the lattice become a set of disconnected $x-z$ zig-zag chains as shown in Fig. \ref{sidebyside}(b). However, when $J_y$ becomes non-zero, finite correlation  develops and reaches the maximum at $\alpha= \pi/4$. For $\alpha$ above $\pi/4$, it starts decreasing and asymptotically reaches to zero at $\alpha = \pi/2$ because this limit corresponds to $J_x=0$ for which the two bonds in the discussion again get disconnected and the lattice become a set of disconnected $y-z$ zig-zag chains as shown in Fig. \ref{sidebyside}(a). Also we observe that in the present context there is no qualitative difference between the four-level and six-level L-Z quench dynamics as expected.

\subsubsection{Correlation function in $(0,0,1)$ direction}

Next we calculate the correlation function for ${\bf r}=(0,0,1)$. The two $z$-bonds are next-neighbor along a $z-x$ zig-zag chain extending in vertical direction. For $\alpha=0$, it is expected to have maximum correlation and should decrease monotonically as we increase $J_y$. The red graph in Figs.~\ref{corr-alpha} (a) and (b) show exactly the expected behavior. The corresponding lattice structure is shown in Fig. \ref{sidebyside} (b). 

\subsubsection{Correlation function in $(1,0,1)$ and $(1,1,1)$ direction}

\begin{figure}
\includegraphics[scale=0.37]{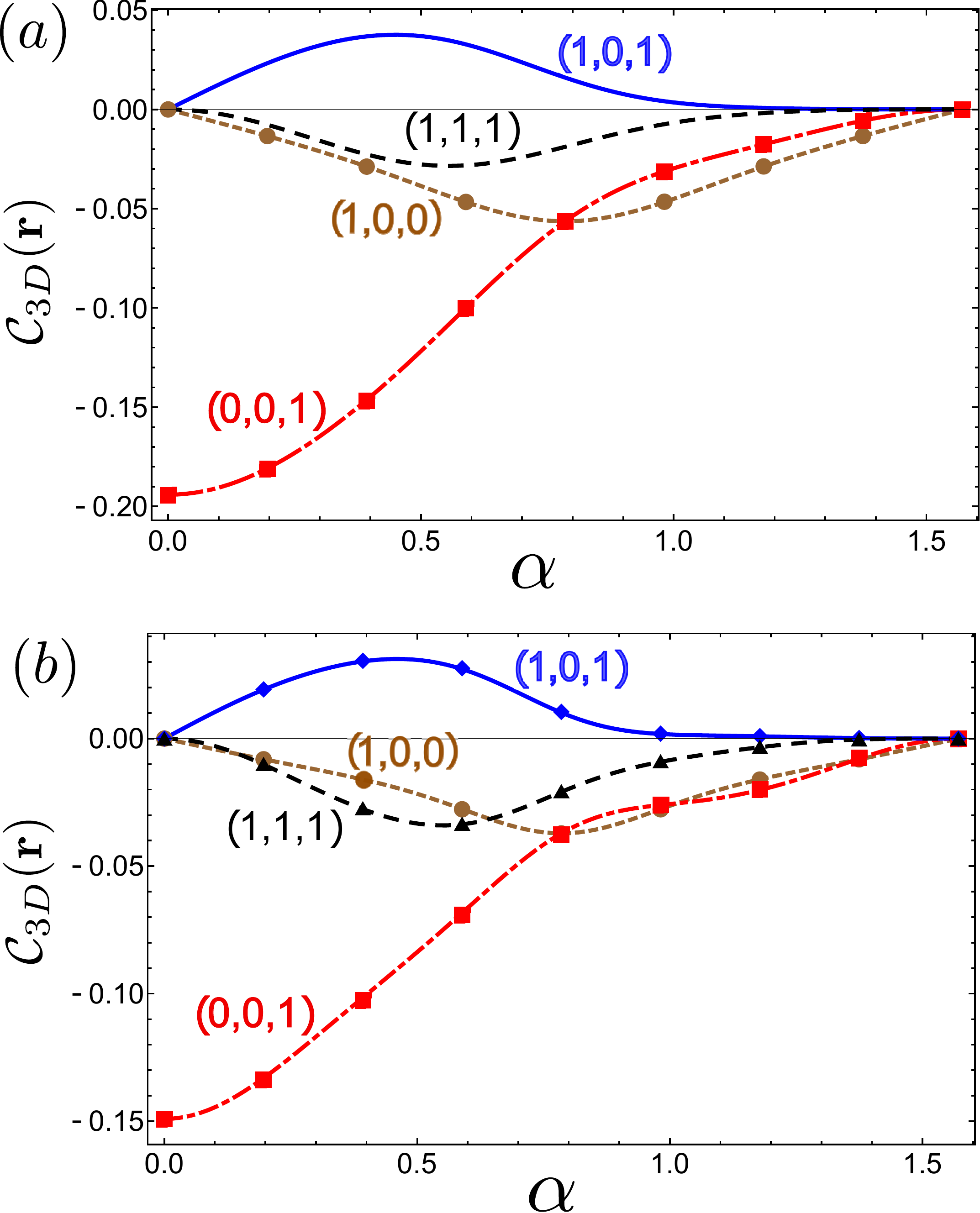}
\caption{\textbf{Variation of correlation function with $\alpha$: }Plot of defect correlation $\langle\mathcal{C}^{3D}(\mathbf{r}) \rangle$ as a function of $\alpha$ where $\alpha = \tan^{-1} (J_y / J_x)$. (a) for the four-level problem, (b) for the six-level problem. The line plots correspond plot to ICA evaluation and the the point plots correspond to the exact numerical calculation of the defect correlation} 
\label{corr-alpha}
\end{figure}

For completeness, we present the correlation functions for ${\bf r}=(1,0,1)$ and ${\bf r}=(1,1,1)$. They are represented by blue and black graphs, respectively in Fig.~\ref{corr-alpha}. We notice that unlike ${\bf r}=(1,0,0)$, the maximum (or minimum) value does not appear for $ \alpha= \pi/4$, though they have zero value for $\alpha=0$ and $\pi/2$. This is attributed to the fact that to connect two $z$-bonds of type-1 in $(1,0,1)$ and $(1,1,1)$, one needs to traverse an unequal number of $x$ and $y$ bonds which results in shifting their respective maxima away from $\alpha= \pi/4$ 

\begin{figure}[h!]
\includegraphics[scale=0.28]{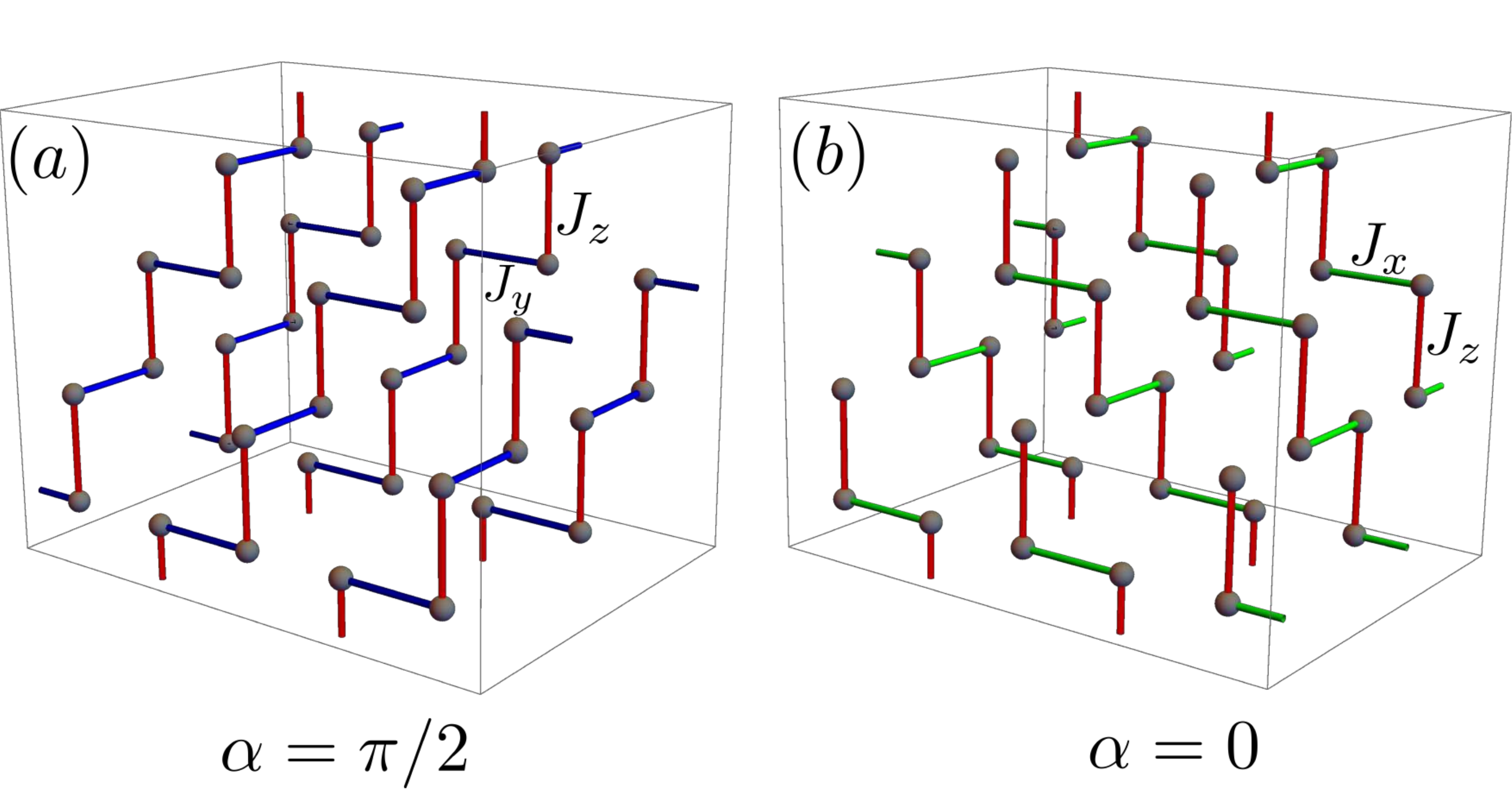}
\caption{\textbf{Lattice structure of 3D Kitaev model}: (a) When $\alpha = \pi / 2$ which corresponds to $J_y, ~ J_z \neq 0$ but $J_x = 0$ (the absent bonds), and (b) when $\alpha = 0 $ which corresponds to $J_x, ~ J_z \neq 0$ but $J_y = 0$ (the absent bonds).}
\label{sidebyside}
\end{figure}
\paragraph*{}
We re-emphasize that the zero value of correlation function in all of the above-mentioned cases can be easily understood by noticing that for $\alpha= \pi/2$, the lattice reduces to disconnected $z-y$ chains as drawn in the right panel of Fig.~\ref{sidebyside} (a). Similarly, at $\alpha=0$ the lattice becomes disconnected $z-x$ chains as depicted in Fig.~\ref{sidebyside} (b). 
\section{Entropy}\label{ent}
In this section, we calculate the von-Neumann entropy of the final quenched state. We note that the initial state is a product of the pure-states and remains so due to the unitary evolution. However, for a given momentum `$k$', the quenched state at any instant is a linear superposition of the instantaneous ground state and the other excited states whose probabilities  exhibit explicit time dependence. It causes the off-diagonal matrix elements of the density matrix to be time-dependent (and oscillatory too) and leads to intrinsic decoherence such that for long quench times, the final quenched state is equivalent to a decohered state \cite{levitov-2006}. It is, therefore, pertinent to calculate the von-Neumann entropy of this quenched states at $t \rightarrow \infty$. The physical reason behind this entropy generation is the underlying non-equilibrium processes in the quench dynamics and subsequent production of defects.
\paragraph*{}
As explained above, for the four-level L-Z problem, the instantaneous state of the system $|\Psi_4 (k, t) \rangle $ can be expressed as $|\Psi_4 (k, t) \rangle = \sum_i \sqrt{A_i (k, t)} |i(k,t) \rangle, ~~i=0,1,2,3$, where $|i(k, t) \rangle$ represents the instantaneous diabatic states(see Fig. \ref{dia-adia-label}) and $A_i(k,t)$ are the respective probabilities.

Keeping in mind that the off-diagonal element of the density matrix constructed from $|\Psi_4 (k, t) \rangle$ does not contribute to the expectation value of any thermodynamic quantity \cite{levitov-2006}, the von-Neumann entropy of the quenched state at $t=+\infty$ is given by,
\begin{equation}
   S= - \int_{HBZ} \frac{d^3 k}{\Omega_{HBZ}} \sum_{i} A_{i}(k, \infty) \ln[A_i(k, \infty)], \label{4en3} 
\end{equation}
where the above equation contains only the time independent diagonal elements of the density matrix. The above equation is evaluated numerically to obtain the von-Neumann entropy of the final decohered state. Moreover, one can write an equivalent expression of $|\Psi_{4}(k, t) \rangle$ within ICA as,
\begin{eqnarray}
 |\Psi_{4} (k, t)\rangle_{ICA} &=& \sqrt{P_{00}} e^{-i E_{0,k} t} |0\rangle+\sqrt{P_{02}} e^{-i E_{2,k} t} |2\rangle \nonumber \\
 && + \sqrt{P_{03}} e^{-i E_{3,k} t} |3\rangle \label{4enica},
\end{eqnarray} 
where the eigenvalues $E_{n,k}$ are essentially corresponding to $t = \pm \infty$ and $P_{ij}$'s are the matrix elements of the transition matrix \eqref{trm}. However, the final density matrix of the system still remains diagonal as the rapidly oscillating off-diagonal terms do not affect any physical quantity ~\cite{levitov-2006,uma-2007}. 
Therefore, the von-Neumann entropy can be evaluated in the quenched state by the following formula,
\begin{equation}\label{Sica}
S^{(ICA)} = -\frac{1}{\Omega_{HBZ}} \int_{HBZ} d^3 k [ \sum_{i}  P_{0i} \ln P_{0i}],~ i=0,2,3.
\end{equation}
  In the limit of very small $\tau$, i.e., $\tau \rightarrow 0$, $P_{00} \rightarrow1 $ and $P_{02}= P_{03} = 0$, the von-Neumann entropy is zero. A procedure similar to \eqref{4en3} can be implemented for the six-level L-Z problem, and the von-Neumann entropy can be computed numerically. However, contrary to the four-level problem, an expression of the quenched state equivalent to \eqref{4enica} can not be obtained for the six-level L-Z problem due to the appearance of the dynamic phases in some of the probability amplitudes (see Appendix \ref{ta1}).
\begin{figure}[h]
\includegraphics[scale=0.65]{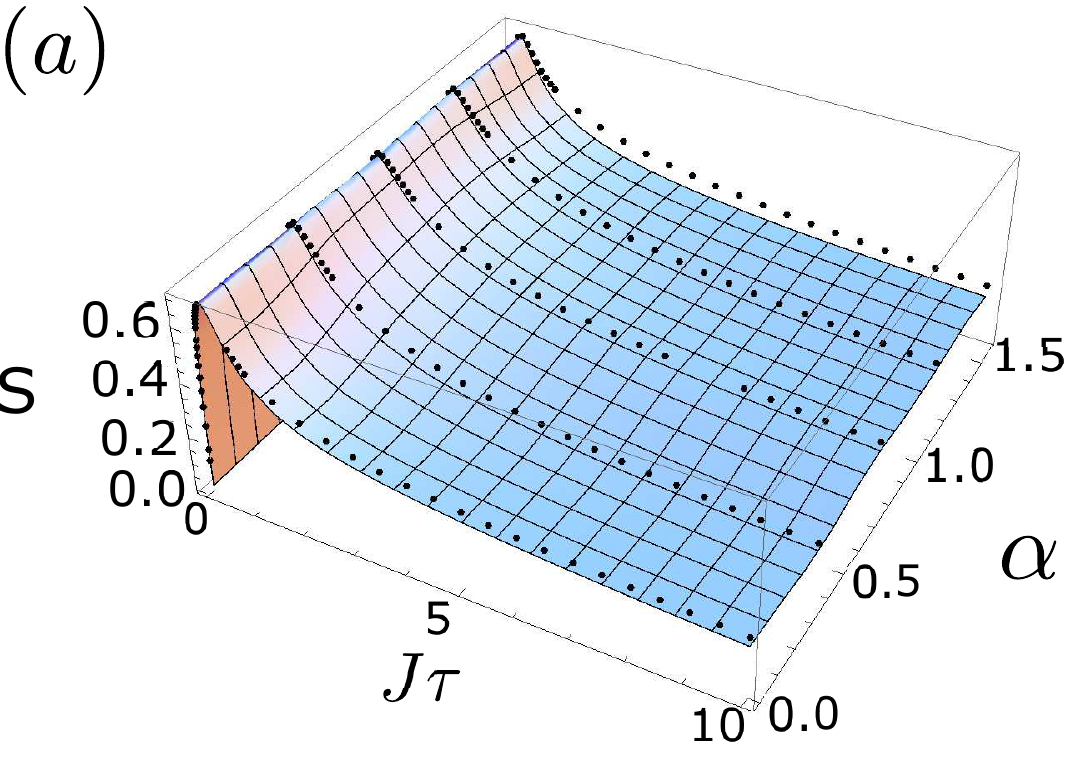}
\includegraphics[scale=0.4]{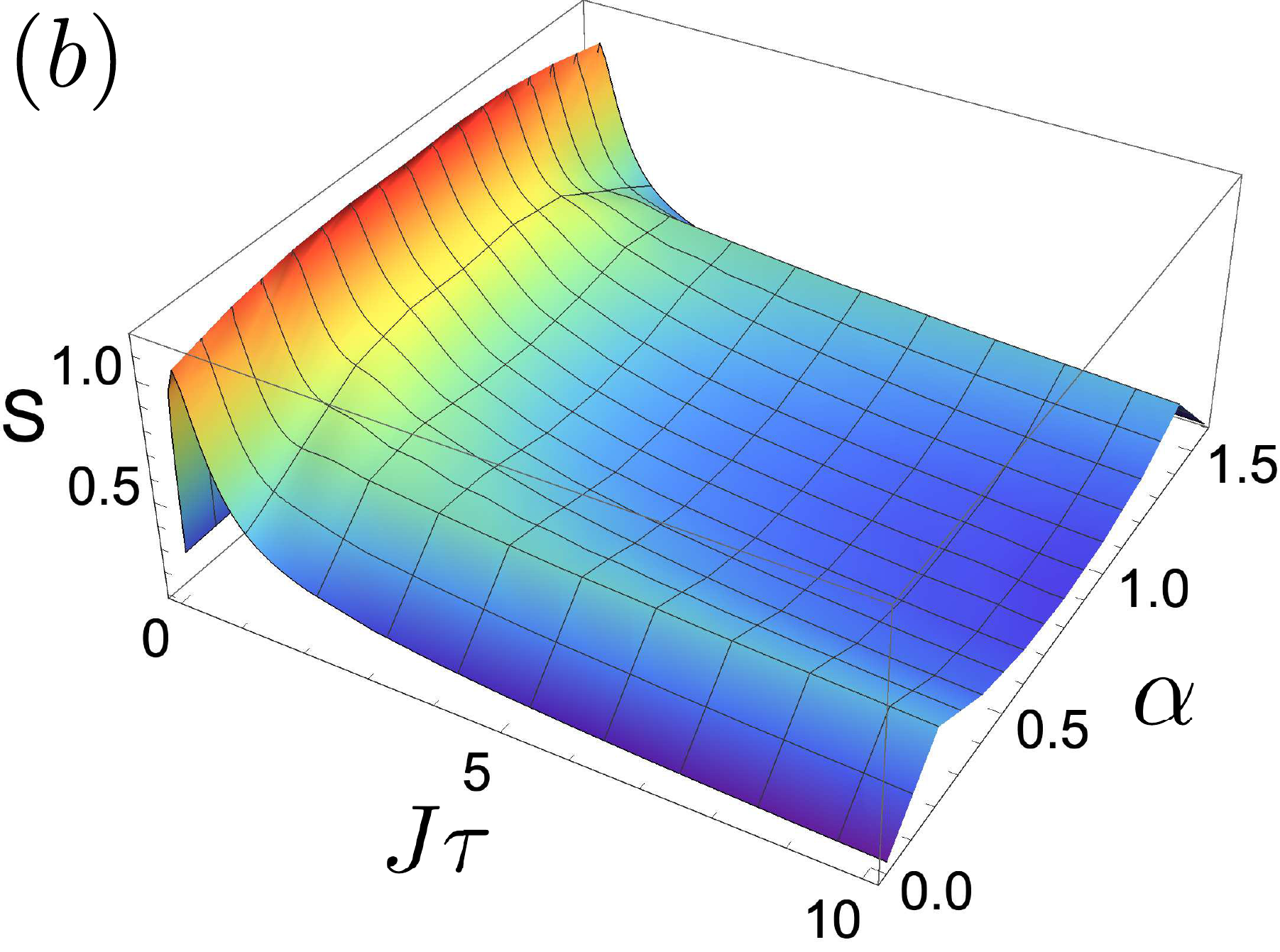}
\caption{\textbf{von-Neumann entropy:} Plot of von-Neumann entropy `$S$' as a function of $J\tau$ and $\alpha$. (a) The plot corresponds to the four-level problem where black dots are obtained from exact numerical calculations of \eqref{4en3}, and the three-dimensional plot is obtained from the ICA (evaluating \eqref{Sica}). (b) The plot corresponds to the six-level problem obtained by exact numerical calculation of \eqref{4en3} but, with $i=0 ~- ~5$. In this case the reason for not evaluating the $S^{(ICA)}$ is explained in the text.}
\label{3dent}
\end{figure}
\begin{figure}[h]
\includegraphics[scale=0.72]{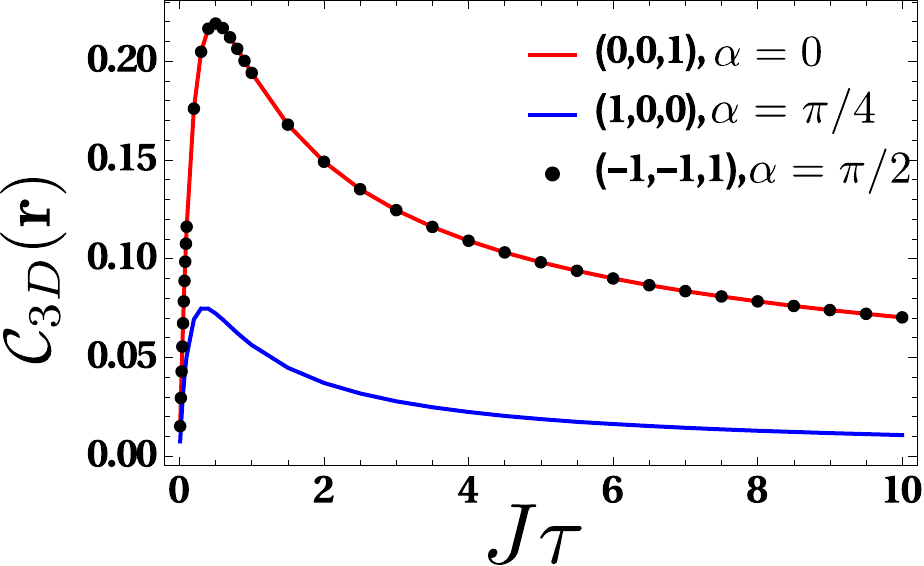}
\includegraphics[scale=0.30]{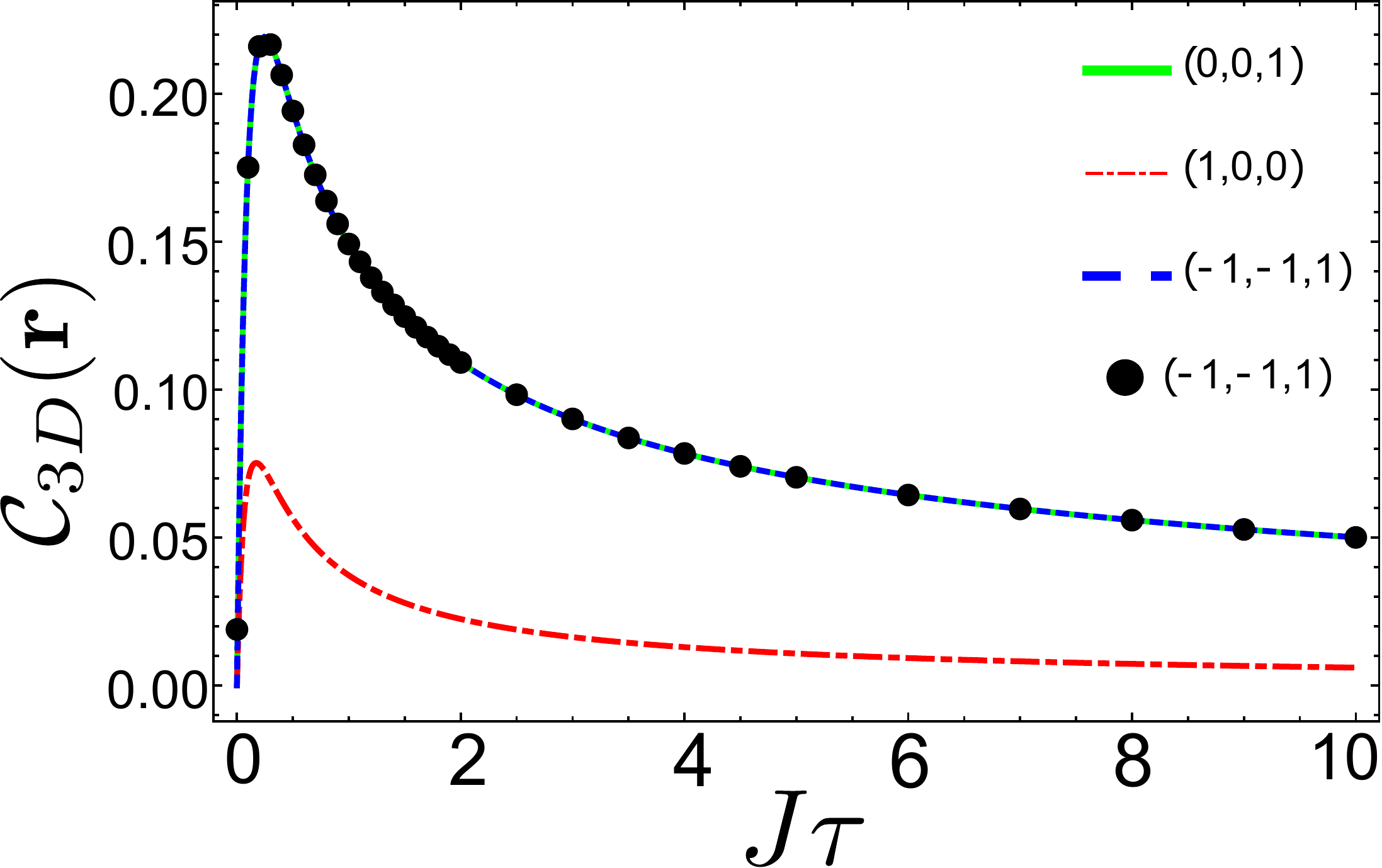}
\caption{\textbf{Variation of correlation functions with $J\tau$}: Plot of correlation function (as defined earlier) as a function of $J\tau$ and for a few representative values of $\alpha$ and direction. Upper panel: four-level problem. Lower panel: six-level problem. In both plots, the black dots are numerically obtained and the solid-lines are obtained from ICA. We see that though maximum of correlation function explains the maximum of entropy  given in Fig.~\ref{3dent}, it can not explain why the maximum of the entropy is almost constant for all $\alpha$. (1,0,0) and (0,1,0) directions correlation are identical for symmetry reasons. Here maximum correlation function obtained for a given $\alpha$ has been plotted.}
\label{corr-jtau}
\end{figure}

The von-Neumann entropy of the initial state is zero because in this case $P_{00}=1$ (see \eqref{Sica}). The non-equilibrium process caused by quenching is expected to generate finite entropy by inducing transitions to the other states such that $P_{00}<1$. The von-Neumann entropy is plotted as a function of $J\tau$ and $\alpha$ in Figs.~\ref{3dent}(a) and \ref{3dent}(b) for the four-level and six-level problems, respectively. We notice from Fig. \ref{3dent} that for a given $\alpha$, the entropy reaches the maximum at a certain value of $J\tau$, and then decreases monotonically and asymptotically saturates to a finite value. This finite value appearing at $t \rightarrow + \infty$ is due to the fact that the final state is a linear superposition of the ground state of $t=+\infty$, and the defect states.

To obtain a more physical insight to this final  decohered state originating from the quenching, we plot the correlation function as a function of $J\tau$ for some representative values of $\alpha$ in Fig.~\ref{corr-jtau}. The correlation function exhibits the form, $\frac{1}{\tau} e^{- A \tau}$ derived in \eqref{cortau}, as a function of quench time $J\tau$. Comparing Figs. \ref{3dent} and \ref{corr-jtau}, we see that for a given $\alpha$, the entropy behaves the same as the correlation function as a function of $J\tau$. It indicates that the quenching generates, through the defect production, the correlation between the Majorana fermions (and therefore, spins) beyond the nearest neighbors, which manifests itself in the increased entropy of the final state. Also, the similar dependence of the entropy (of the final state) and the correlation function as a function of $J\tau$ conform to the fact that the quenched state is an inherently decohered state whose density matrix can be represented by the time independent diagonal elements. These diagonal elements, in effect, determine the expectation values of any thermodynamic quantities (correlation function in this case)\cite{levitov-2006}.

\section{Conclusions and discussions}\label{conc}
To summarize, we have studied the quench dynamics and defect production in the spin- 1/2 3D Kitaev model within the set-up of a multi-level L-Z problem. By mapping the spin Hamiltonian to a quadratic Hamiltonian of Majorana fermions hopping in the background of a uniform static $\mathbb{Z}_2$ gauge field corresponding to each bond, we have obtain a Hamiltonian with $4\times4$ block diagonal form in the momentum space. We have constructed a four-level and a six-level L-Z problem by filling up the lowest negative energy state and both the negative energy states respectively. Our analytical calculations using the ICA agree with the corresponding exact numerical calculations. Our major findings are the following:
\begin{itemize}
    \item[1.] In the limit of very slow quench, viz., $\tau >> 1$ the defect density $n_d$ scales as $\tau^{-1}$ with the quench time $\tau$ and satisfies the general scaling law $n_d = \tau^{-\frac{m \nu}{z\nu + 1}}$, with $m=2$, and $\nu = z =1$, where $\nu$ is the critical exponent of the correlation length, and $z$ is the dynamic critical exponent. Moreover, in the $\tau >> 1$ limit for a fixed $\tau$ the defect density is found to be maximum at $\alpha = 0$ and $\pi /2$, and at $\pi / 4$ exhibits a local maximum. 
    \item[2.] In the limit of very slow quench the defect correlation function is found to scale as $\tau^{-1} e^{-A/\tau}$ with the quench time $\tau$. Furthermore, the correlation function is found to be spatially symmetric in the $x-y$-plane (or, equivalently, in the $\mathbf{a}_1 - \mathbf{a}_2$ plane), and asymmetric in the other two planes for $J_x = J_y$. However, in general, the defect correlation is an-isotropic for $J_x \neq J_y$.
    \item[3.] We have evaluated the residual   entropy in the final state after quenching as a function of the system-parameter. The dependence of the entropy on the quench rate is qualitatively the same as that of the correlation function.
\end{itemize}
In the following, we discuss all the above mentioned findings one by one.
\paragraph*{}
The scaling behavior of the defect density obtained in the 3D Kitaev model considered hare is a direct consequence of the way the system enters into the non-adiabatic regime. The non-adiabatic condition, governed by $|\Delta_{1k}|^2 = 0$, where $\Delta_{1k}$ is the relevant coupling between the levels determining the defect density, 
does not ensure the spectrum of the system to be gapless. The fact that the defect density satisfies the general scaling law $n_d = \tau^{-\frac{m \nu}{z\nu + 1}}$, proposed in Ref. \cite{Pol2, KS2}, indicates a deeper connection between the gapless condition for the energy spectrum and the non-adiabatic condition. For the present model, the parameter $\Delta_{1k}$ appears within the condition for the spectrum to be gapless, but the zeros of $\Delta_{1k}$ do not ensure the gaplessness of the spectrum. Incidentally, the zeros of $\Delta_{1k}$ and the spectrum  both constitute their own set of $d-m =1$ dimensional critical lines which ensure the general scaling law being satisfied. This is unlike the 2D Kitaev model where the gapless condition itself determines the non-adiabaticity \cite{KS2}. We further speculate that the departure of the scaling law from the above mentioned behavior may be possible in a multi-level L-Z problem, if the relevant non-adiabatic condition exhibits a different dimensional dependence than the condition for the energy spectrum to be gapless. Finding such a possibility or lack of it, we believe, is a definitive direction of theoretical investigation in the future.
\paragraph*{} The dependence of the defect density on $\alpha$ originates form the nature of the effective-DOS corresponding to the function $\Delta_{1k}$ in the gapless phase as shown in Fig. \ref{defect_alpha}. Physically the effective-DOS quantifies the number of $\mathbf{k}$-states satisfying the non-adiabatic condition. When the quenching process takes the system across the gapless phase the number of $\mathbf{k}$-states satisfying the non-adiabatic condition is the maximum at $\alpha = 0$ and $\pi /2$. This leads to a larger defect formation at these values of $\alpha$. On the other hand, the number of $\mathbf{k}$-states satisfying the non-adiabatic condition at $\alpha = \pi /4$ is locally maximum, making the defect density locally maximum as well.
\paragraph*{}
Moreover, when $J_x = J_y$ the lattice retains its full 3D structure and the nearest-neighbor dimers are placed along the $\mathbf{a}_1$ and $\mathbf{a}_2$ directions. From Fig. \ref{3d-phs-d} it is easy to recognize that when $J_x = J_y$ a dimer finds an equal number of nearest-neighbor dimers in both $\mathbf{a}_1$ and $\mathbf{a}_2$ directions. This makes the defect correlation in the $\mathbf{a}_1 - \mathbf{a}_2$ plane to become symmetric with respect to the $|\mathbf{a}_1| = |\mathbf{a}_2|$ line. For other planes such a condition is not satisfied. It turns out that when either $J_x$ or $J_y$ vanishes, the lattice becomes a set of decoupled chains (see Fig. \ref{sidebyside}) and only those correlations survive for which a dimer can find another dimer in the same chain. 

\paragraph*{}
Starting from a trivial product state of ferromagnetic dimers the quenching process takes the system to a final state  whose density matrix can be approximated as a suitable reduced density matrix exhibiting a finite value of the von-Neumann entropy. Most certainly, the final  state contains the ground state of the Hamiltonian at $t=\infty$, and the other excited states. The post quench (at $t=\infty$) ground state of the Hamiltonian exhibits a structure where anti-ferromagnetic dimers are sitting on each $z$-links. In this regard, it is worthwhile to point out that our definition of the defect density quantifies the ferromagnetic component of the final  state. The maxima of the correlation functions and the   entropy are reached around same values of $J \tau$. This indicates that the quenching process gives rise to the generation of a non-local correlation which is further manifested in the finite values   entropy. 

\paragraph*{}
There exists a class of 3D materials, such as $\beta-$ and $\gamma- \text{Li}_2\text{IrO}_3$, containing Kitaev interactions as the effective low energy description \cite{winter}. These materials offer a possibility to experimentally realize the quench dynamics in the 3D Kitaev model studied here. However, in these materials apart from the Kitaev interactions other interactions such as, Heisenberg interaction, asymmetric interaction of the type $S^{x}S^{y}$, or even Dzyaloshinskii-Moriya interaction are present which destroy the exact solvability of the system. We expect that the external pressure or magnetic field may become useful in realizing a similar quench dynamics given the fact that these parameters have been useful to realize the much-coveted spin-liquid phase in the Kitaev materials \cite{Taka,breznay,veiga}. In particular, it has been pointed out that $x-x,~~y-y$ or $z-z$ type interactions can be changed by applying pressure \cite{veiga}. Therefore, one can, in principle, achieve the quench protocol we have considered, viz., the variation of the strength of $z-z$ interaction. We also believe that an alternative way to realize our quench study, as well as quench studies in other multi-level L-Z problem, would be in the optical lattices as envisaged in a recent experiment \cite{keesling}.
\section*{Acknowledgements}
S.M. acknowledges the  earlier collaboration with Anirban Dutta with whom the project actually started but could not be continued. S.S. acknowledges the partial financial support provided through a Kreitman postdoctoral fellowship and the Israel Science Foundation (ISF) Grant No.~1360/17, and Sayak Ray for critically reading the manuscript.

\begin{appendix}

\section{Ground states of the single and the two-particle sector}\label{tb}
We have performed a unitary transformation to diagonalize the upper-left and lower-right $2\times2$ blocks of \eqref{hkmat}. As a result, we get the following relationship between the old $\Psi_{k}$ fermions (defined in \eqref{hkmat}) and new $\eta_{i,k}$ fermions:
\begin{eqnarray}
\label{ukpk}\begin{pmatrix} \psi_k \\
\phi_k \end{pmatrix}
 &=&  u_k \begin{pmatrix} \eta_{3k} \\
\eta_{2k} \end{pmatrix},~~\begin{pmatrix} \psi^{\dagger}_{-k} \\
\phi^{\dagger}_{-k} \end{pmatrix} =  u_k \begin{pmatrix} \eta_{0k} \\
\eta_{1k} \end{pmatrix},
\end{eqnarray}
where $u_k$ is the $2 \times 2$ unitary matrix used in the above-mentioned diagonalization, and `$k$' has been defined only over the HBZ. The explicit expression for $u_k$ is given by,
\begin{eqnarray}
&&u_k= \begin{pmatrix}-e^{-i \theta_{2k}} & e^{-i \theta_{2k}} \\
1 &1    \end{pmatrix},~~~ e^{i \theta_{2k}}= \frac{\Delta_{2k}}{|\Delta_{2k}|}.
\end{eqnarray}

The initial state for the four-level problem, which we denote by $| \mathcal{I} \rangle $, is obtained by filling only the $\eta_{0k}$ of the vacuum of $\eta_{ik}$ ($| \tilde{\emptyset} \rangle $). Similarly, the ground state $| \mathcal{G} \rangle $ for the six-level problem is obtained by filling $\eta_{0k}$ and $\eta_{1k}$. We then find the following relations,
\begin{eqnarray}
|\tilde{\emptyset} \rangle &&= \eta_{0k} \eta_{1k}  | \emptyset \rangle, \\
| \mathcal{I} \rangle && = \eta^{\dagger}_{0k} | \tilde{\emptyset} \rangle,~~~~| \mathcal{G} \rangle =\eta^{\dagger}_{1k} \eta^{\dagger}_{0k} | \tilde{\emptyset} \rangle .
\end{eqnarray}
In the above, $| \emptyset \rangle$ is the original vacuum of the fermions representing a zero-fermion state, and we used the fact that $| \emptyset \rangle$ is already a vacuum for $\eta_{2k}$ and $\eta_{3k}$.

All four singly-occupied states corresponding to the four-level problem are given by, $|0\rangle = \eta_{0k}^{\dagger} |\tilde{\emptyset} \rangle$, $|1\rangle = \eta_{1k}^{\dagger} |\tilde{\emptyset} \rangle$ , $|2\rangle = \eta_{2k}^{\dagger} |\tilde{\emptyset} \rangle$, and $|3\rangle = \eta_{3k}^{\dagger} |\tilde{\emptyset} \rangle$. With these notations, we have defined $\Phi_{4,k}$ in \eqref{4state-coupled}. Similarly, for the six-level problem, we enumerate all the six doubly occupied states in terms of the single particle states and their respective energies as follows:

 \begin{eqnarray}
 &&|\tilde{0} \rangle= \eta^{\dagger}_{0k} \eta^{\dagger}_{1k} | \tilde{\emptyset} \rangle,~~ \tilde{\epsilon}_0= 4 J_3, \\
 && | \tilde{1} \rangle= \eta^{\dagger}_{0k} \eta^{\dagger}_{2k} | \tilde{\emptyset} \rangle,~~\tilde{\epsilon}_1=0,\\
 &&  | \tilde{2} \rangle= \eta^{\dagger}_{1k} \eta^{\dagger}_{2k} | \tilde{\emptyset} \rangle,~~\tilde{\epsilon}_2=-2 |\Delta_2|,\\
 &&|\tilde{3} \rangle= \eta^{\dagger}_{0k} \eta^{\dagger}_{3k} | \tilde{\emptyset} \rangle,~~\tilde{\epsilon}_4=2 |\Delta_2|,\\
 &&|\tilde{4} \rangle= \eta^{\dagger}_{1k} \eta^{\dagger}_{3k} | \tilde{\emptyset} \rangle,~~ \tilde{\epsilon}_3=0, \\
 &&  | \tilde{5} \rangle= \eta^{\dagger}_{2k} \eta^{\dagger}_{3k}| \tilde{\emptyset} \rangle,~~\tilde{\epsilon}_5=-4 J_3.
 \end{eqnarray}
 Within this notation, we have defined, in the Schrodinger equation governing the time evolution of the system, $\Phi^{\dagger}_{6,k}= ( |\tilde{0} \rangle, |\tilde{1} \rangle, |\tilde{2} \rangle, |\tilde{3} \rangle,|\tilde{4} \rangle, |\tilde{5} \rangle)$.
 
\section{Transition amplitudes: four-level problem}\label{ta}
To calculate the transition amplitudes, we first consider all possible semi-classical paths corresponding to Fig. \ref{dia-adia-label}(a). As mentioned in section \ref{mod}, the ground state at $t=-\infty$ is $|0\rangle$. First we note that there are four crossing points, viz., 
\begin{eqnarray}\label{cros_p}
|0\rangle &\rightarrow & |3\rangle:\; \text{with coupling constant}:\, g_{k}, \nonumber \\
|3\rangle &\rightarrow & |0\rangle:\; \text{with coupling constant}:\, -g_{k}, \nonumber \\
|1\rangle &\rightarrow & |2\rangle:\; \text{with coupling constant}:\, -g_{k}, \nonumber \\
|2\rangle &\rightarrow & |1\rangle:\; \text{with coupling constant}:\, g_{k}, \nonumber \\
|3\rangle &\rightarrow & |1\rangle:\; \text{with coupling constant}:\, -\gamma_{k}, \nonumber \\
|0\rangle &\rightarrow & |2\rangle:\; \text{with coupling constant}:\, \gamma_{k}, 
\end{eqnarray}
We apply the LZ formula by using the above coupling constants corresponding to the respective crossing points. We note that whenever a state crosses another state with coupling constant $g_k$ (as well as $-g_k$), the probability amplitude that the system remains in the same state is given by $u_{g} = \exp \left(- \frac{\pi |g_k|^2 (J\tau)}{4 J^2} \right)$. Therefore, the probability amplitude that the system makes a transition is $i \xi\sqrt{1 - (u_{g})^2}$, where $\xi =1$ when the sign of the coupling constant is positive and $\xi = -1$, when negative. These expressions for the crossing points with coupling $\gamma_k$ are same except $|g_k|$ shall be replaced by $|\gamma_k|$. In the following, let us write down the transition amplitudes $S_{ij}$ for all the semi-classically allowed paths. In writing such paths we need to consider the effect of the dynamic phase $e^{i \phi_{d}^{ij}(\nu)}$  in a given path $\nu$ which is given by,
\begin{equation}
\phi_{d}^{ij} (\nu) = \int_{-\infty}^{+\infty} dt \epsilon_{\nu k}(t); \, \text{with}\,0,\,1,\,2,\,3,
\end{equation}
where $\epsilon_{\nu k}(t)$ are the diabatic energy levels participating in the trajectory. This dynamic phase is nothing but the area under the trajectory $\nu$ under consideration. All the transition amplitudes corresponding to the transitions from the state $|0\rangle$ to the rest of the states are given by, 
\begin{eqnarray}
|0\rangle &\rightarrow & |0\rangle: S_{00}=u_{g} u_{\gamma}, \nonumber \\
|0\rangle &\rightarrow & |1\rangle: S_{01}= 0,\;\text{not an allowed transition}, \nonumber \\ 
|0\rangle &\rightarrow & |2\rangle: S_{02}=i \sqrt{1- u_{\gamma}^2}, \nonumber \\
|0\rangle &\rightarrow & |3\rangle: S_{03}=i u_{\gamma} \sqrt{1- u_g^2}.
\end{eqnarray}
\begin{figure*}
\subfloat[]{ \includegraphics[width=.4\linewidth]{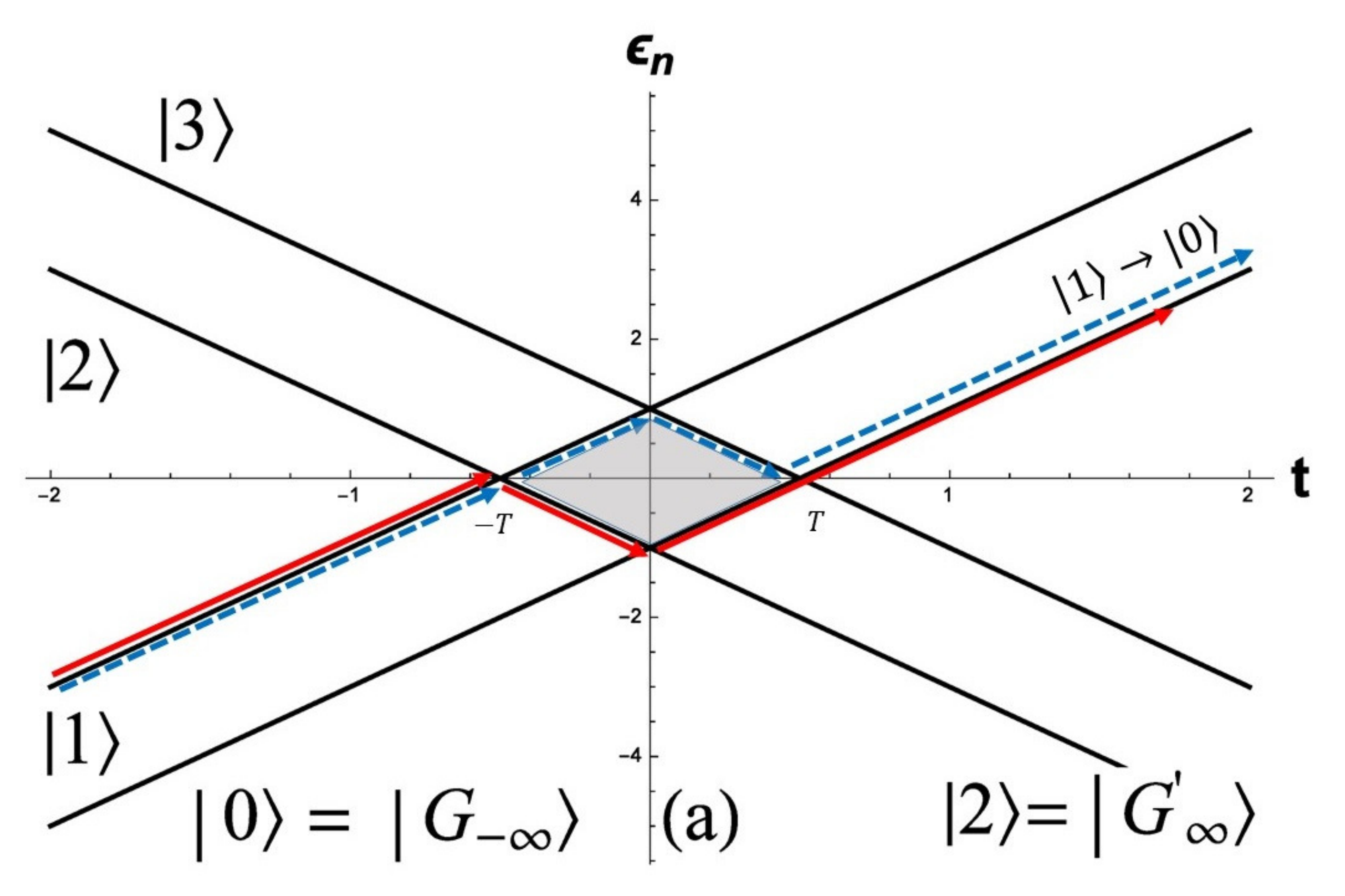}}
\subfloat[]{ \includegraphics[width=.4\linewidth]{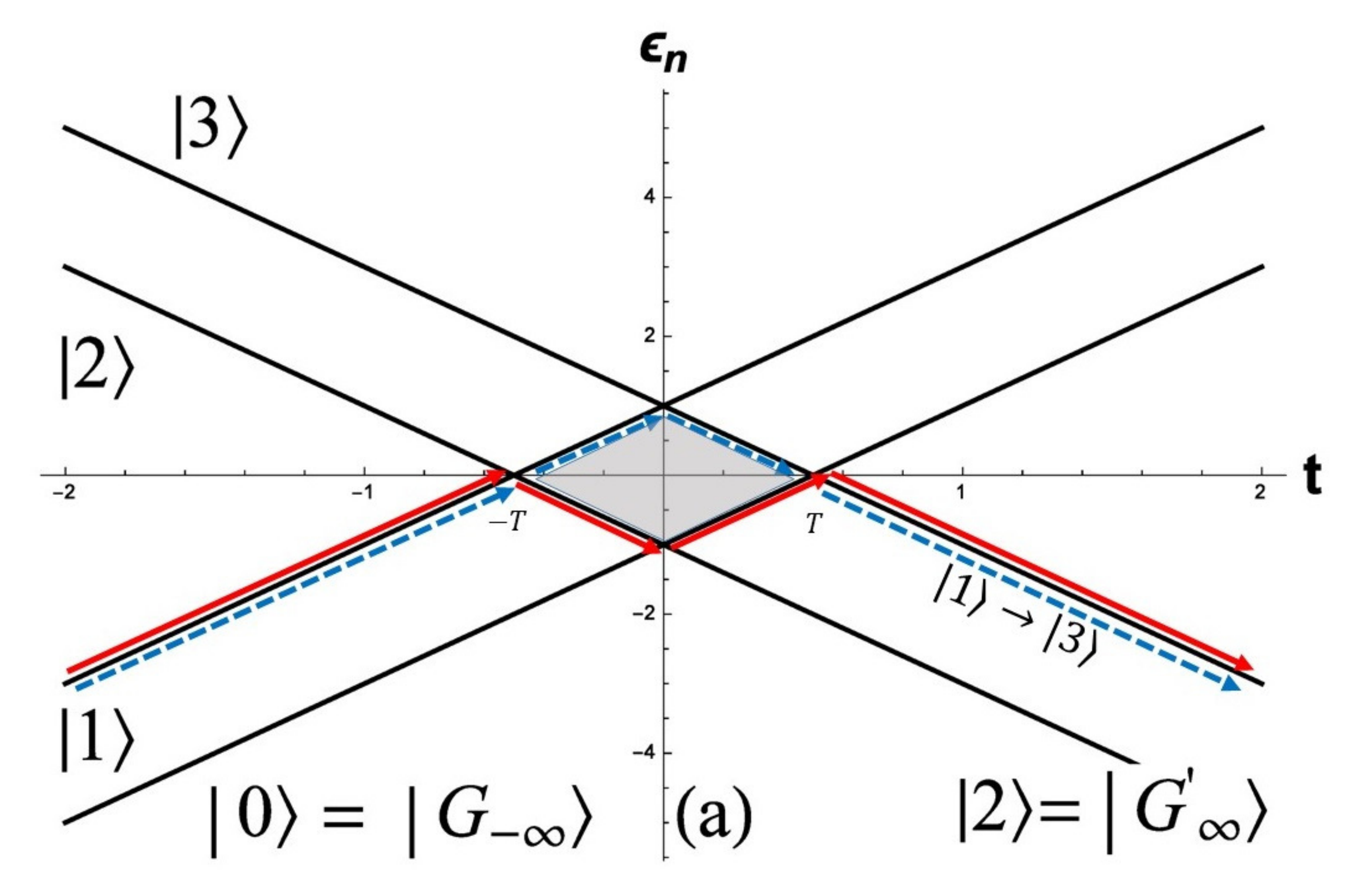}}\\
\subfloat[]{\includegraphics[width=.4\linewidth]{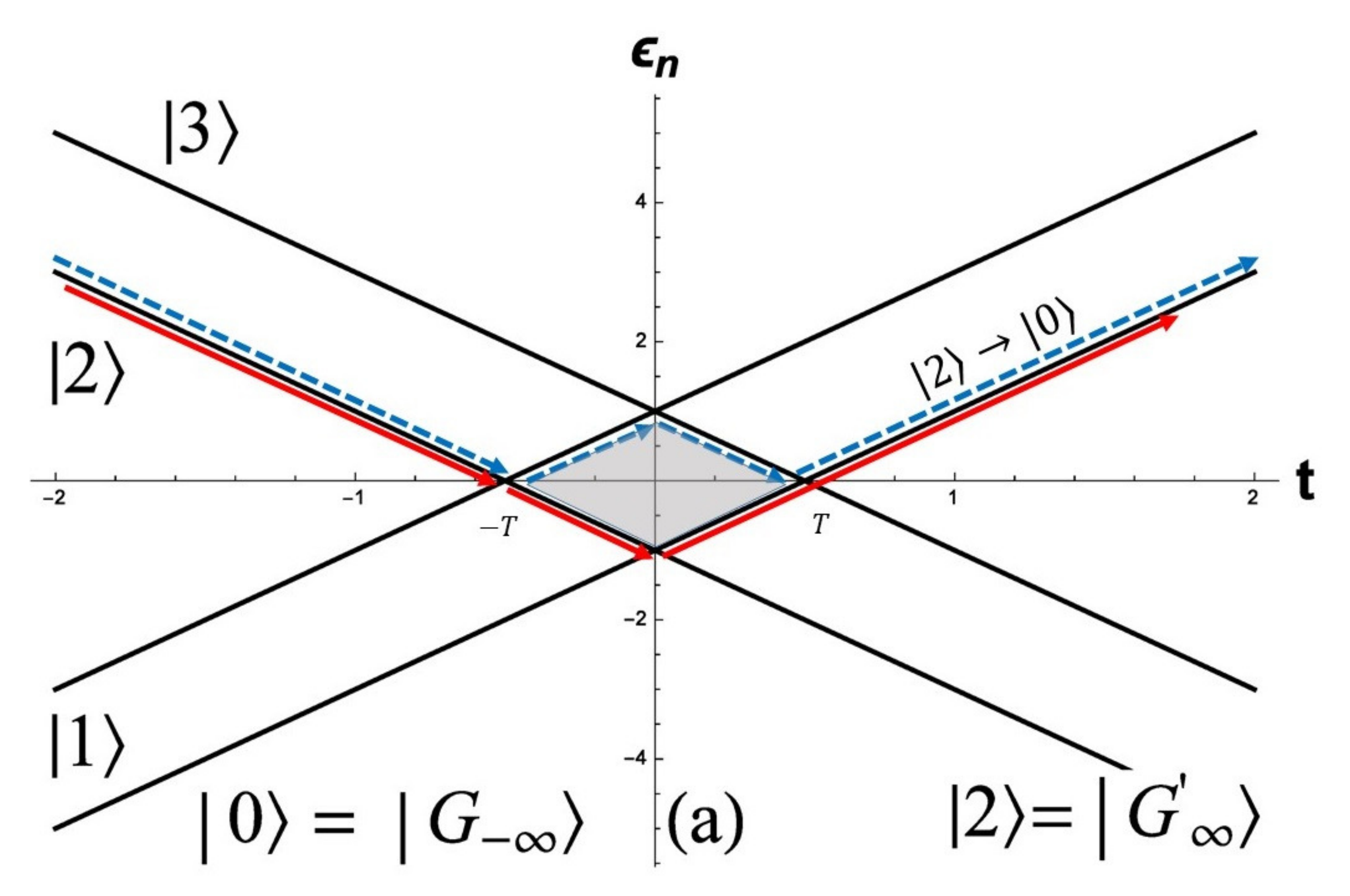}}
 \subfloat[]{\includegraphics[width=.4\linewidth]{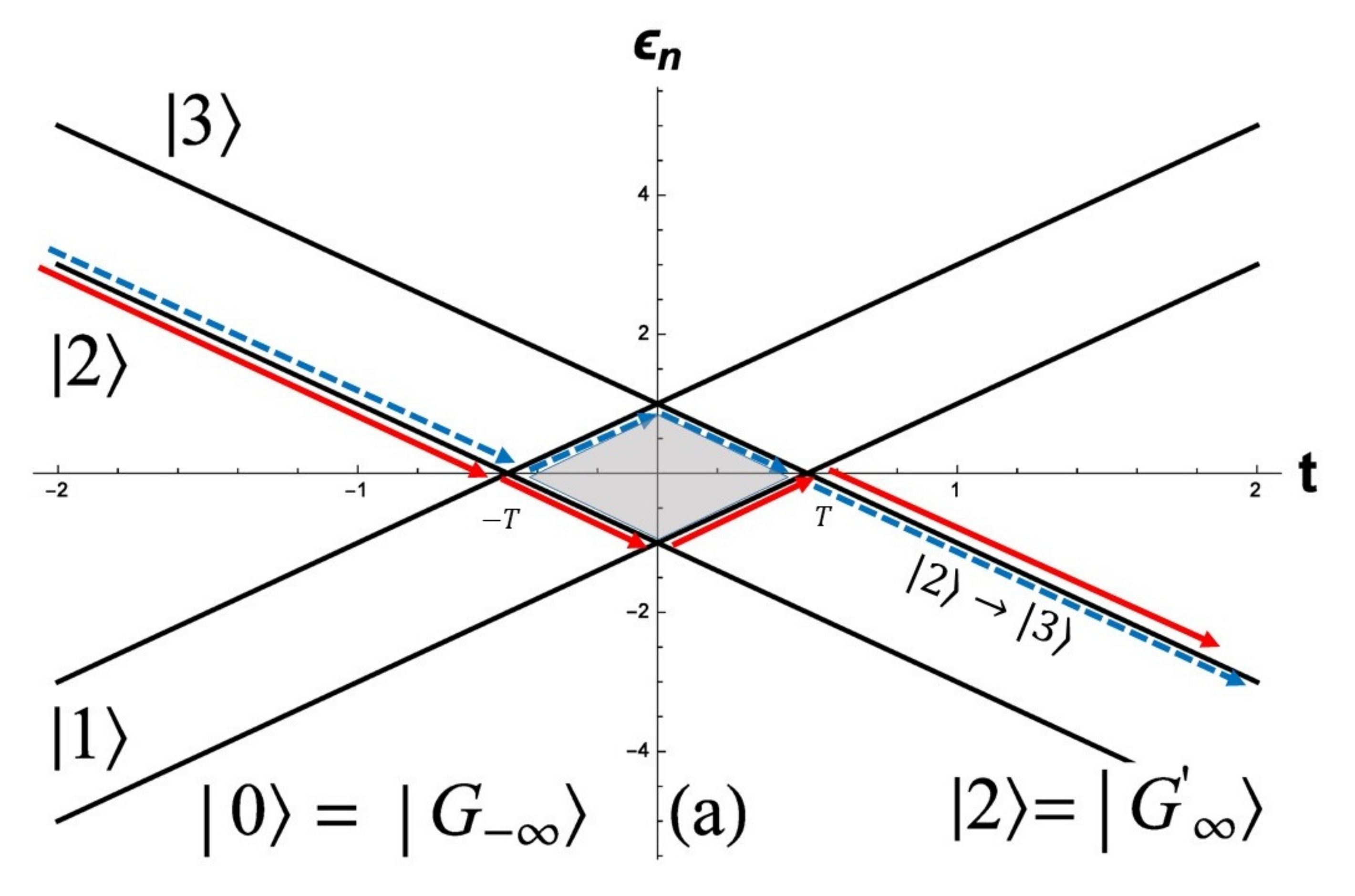}}
\caption{\textbf{Diabatic level diagram and the dynamic phase difference:} Diabatic levels corresponding to the Hamiltonian \ref{4LZ}; (a) The semi-classical trajectories showing the possible paths along which a transition from the state $|1\rangle$ to the state $|0\rangle$ can take place. Similar trajectories corresponding to the transitions from (b) the state $|1\rangle$ to the state $|3\rangle$, (c) the state $|2\rangle$ to the state $|0\rangle$, (d) the state $|2\rangle$ to the state $|3\rangle$ are shown. The blue dashed and red solid trajectories shall henceforth be called trajectories (2) and (1), respectively. The gray shaded region is the dynamic phase difference between the said trajectories.}
\label{1-to-4}
\end{figure*}
In the above equation the dynamic phases in the respective trajectories get canceled when the square of the modulus of the transition amplitudes are taken. Therefore we have dropped the term $e^{i \phi_{d}^{0j}(\nu)}$ corresponding to the dynamic phase. Similarly, all the transition amplitudes corresponding to the transitions from $|1\rangle$ to the rest of the states can be written down. Here, while considering all the semi-classical trajectories, we need to consider the effects of the dynamic phases on the transition amplitude in the corresponding trajectories,
\begin{eqnarray}
|1\rangle &\rightarrow & |1\rangle: S_{11}=u_{g} u_{\gamma}, \nonumber \\
|1\rangle &\rightarrow & |0\rangle: S_{10}=u_{g} \sqrt{(1-u_{g}^{2})(1-u_{\gamma}^{2})} \left[e^{i \phi_{d}^{10}(1)}-e^{i \phi_{d}^{10}(2)}\right], \nonumber \\
|1\rangle &\rightarrow & |3\rangle: S_{13}=i\sqrt{1-u_{\gamma}^{2}}\left[(1-u_{g}^{2})e^{i \phi_{d}^{13}(1)} - u_{g}^{2} e^{i \phi_{d}^{13}(2)}\right], \nonumber \\
|1\rangle &\rightarrow & |2\rangle: S_{12}=i \left(\sqrt{1- u_{g}^2}\right) u_{\gamma},
\end{eqnarray}
where in the above equation $e^{i \phi_{d}^{ij}(1)}$ and $e^{i \phi_{d}^{ij}(2)}$ are the dynamic phases corresponding to the trajectories (1) and (2), respectively, and the phase difference is given by, $\Delta \phi_{d}(2,1) = \phi_{d}^{ij}(2)-\phi_{d}^{ij}(1) = 4 J_z T^2 + |\Delta_{2k}| T $, which is the area in the $\epsilon - t$ plane bounded by both the trajectories. $\pm T$ are the points on the $t$-axis where the diabatic levels cross each other and the time axis too. It is easy to recognize from Fig. \ref{1-to-4} that the phase difference is the same for all the four transition amplitudes where the dynamic phase plays its role. Likewise, all  the transition amplitudes corresponding to the transitions from $|2\rangle$ are given by, 
\begin{eqnarray}
|2\rangle &\rightarrow & |2\rangle: S_{22}=u_{g} u_{\gamma} \nonumber \\
|2\rangle &\rightarrow & |3\rangle: S_{23}= -S_{10}\, (\text{from symmetry})\nonumber \\
|2\rangle &\rightarrow & |0\rangle: S_{20}=-S_{13} \, (\text{from symmetry}) \nonumber \\
|2\rangle &\rightarrow & |1\rangle: S_{21}=i \left(\sqrt{1- u_{g}^2}\right) u_{\gamma}.
\end{eqnarray}
Lastly, all transition amplitudes corresponding to transitions from the state $|3 \rangle$ are given by,
\begin{eqnarray}
|3\rangle &\rightarrow & |3\rangle: S_{33}=u_{g} u_{\gamma}, \nonumber \\
|3\rangle &\rightarrow & |2\rangle: S_{32}=0,\;\text{not an allowed transition}, \nonumber \\
|3\rangle &\rightarrow & |1\rangle: S_{31}= -i \sqrt{1- u_{\gamma}^2}, \nonumber \\
|3\rangle &\rightarrow & |0\rangle: S_{30}= -i u_{\gamma} \sqrt{1- u_g^2}.
\end{eqnarray}
Therefore, the transition probability matrix is given by,
\begin{eqnarray}\label{trm}
P_{\sarkar{4}} = \begin{pmatrix}
(u_{g} u_{\gamma})^{2} & 0 & (1-u_{\gamma}^{2}) & u_{\gamma}^2 (1-u_{g}^{2}) \\
|S_{10}|^2 & (u_{g} u_{\gamma})^{2} &u_{\gamma}^2 (1-u_{g}^{2}) & |S_{13}|^2  \\
|S_{13}|^2 & u_{\gamma}^2 (1-u_{g}^{2}) & (u_{g} u_{\gamma})^{2} & |S_{10}|^2\\
 u_{\gamma}^2 (1-u_{g}^{2})& (1-u_{\gamma}^{2}) & 0 & (u_{g} u_{\gamma})^{2}
\end{pmatrix}, \nonumber \\
\end{eqnarray}
where the matrix elements of the above matrix is given by, $P_{ij} = S_{ij}^{*} S_{ij}$, and
\begin{eqnarray}
P_{10} = |S_{10}|^2 &=& 4u_{g}^{2}(1-u_{g}^{2})(1-u_{\gamma}^{2}) \sin^{2} \left[\frac{\Delta \phi_{d}(2,1)}{2} \right],\nonumber \\
P_{13} = |S_{13}|^2 &=& \left[1-4u_{g}^{2}(1-u_{g}^{2})\right](1-u_{\gamma}^{2}) \times \nonumber \\ & & \cos^{2} \left[\frac{\Delta \phi_{d}(2,1)}{2} \right].
\end{eqnarray}
The quantity $P^{(4)}_{00}$ used in \eqref{P00} is given by $P^{(4)}_{00} = (u_{g} u_{\gamma})^{2}$.

\section{Transition amplitudes: six-level problem}\label{ta1}
The six-level matrix representation for the 3D Kitaev Model is given by,
\begin{eqnarray}
 \tilde{\mathcal{U}}_{k}= \left(  \begin{array}{cccccc}
 4 J_3 & \gamma_k & - g_k & g_k & - \gamma_k & 0 \\
 \gamma_k & \epsilon & 0& 0& 0& - \gamma_k \\
 -g_k &0& 2 |\Delta_{2k}|&0&0& g_k \\
 g_k&0&0&-2|\Delta_{2k}|&0&-g_k \\
 -\gamma_k&0&0&0&-\epsilon &\gamma_k \\
 0& -\gamma_k & g_k & -g_k& \gamma_k&-4J_3 \end{array} \right)~~
 \end{eqnarray}
with $J_3 = J t/ \tau$, where the degeneracy removal term $\pm \epsilon$ has been used for the horizontal levels at zero energy. This is perhaps the only way to apply ICA in a system exhibiting more than one degenerate horizontal energy levels passing through zero energy. However, to take the limit $\epsilon \rightarrow 0$ is mandatory at the end of the calculation, and in this limit $T_1$ also vanishes. 
The couplings at the crossings are given by,
\begin{eqnarray}
 |\tilde{0} \rangle \longleftrightarrow |\tilde{2} \rangle & : & g_k  ;\text{at}\; t= -T_2, \nonumber \\
 |\tilde{0} \rangle \longleftrightarrow |\tilde{1} \rangle & : & -\gamma_k ;\text{at}\; t= -T_1, \nonumber \\
 |\tilde{0} \rangle \longleftrightarrow |\tilde{4} \rangle & : & \gamma_k ;\text{at}\; t= T_1, \nonumber \\
 |\tilde{0} \rangle \longleftrightarrow |\tilde{3} \rangle & : & -g_k ;\text{at}\; t= T_2, \nonumber \\
 |\tilde{5} \rangle \longleftrightarrow |\tilde{2} \rangle & : & -g_k  ;\text{at}\; t= T_2, \nonumber \\
 |\tilde{5} \rangle \longleftrightarrow |\tilde{1} \rangle & : & \gamma_k ;\text{at}\; t= T_1, \nonumber \\
 |\tilde{5} \rangle \longleftrightarrow |\tilde{4} \rangle & : & -\gamma_k ;\text{at}\; t= -T_1,  \nonumber \\
 |\tilde{5} \rangle \longleftrightarrow |\tilde{3} \rangle & : & g_k ;\text{at}\; t= -T_2.
\end{eqnarray}
\begin{figure}
    \centering
    \includegraphics[scale=0.50]{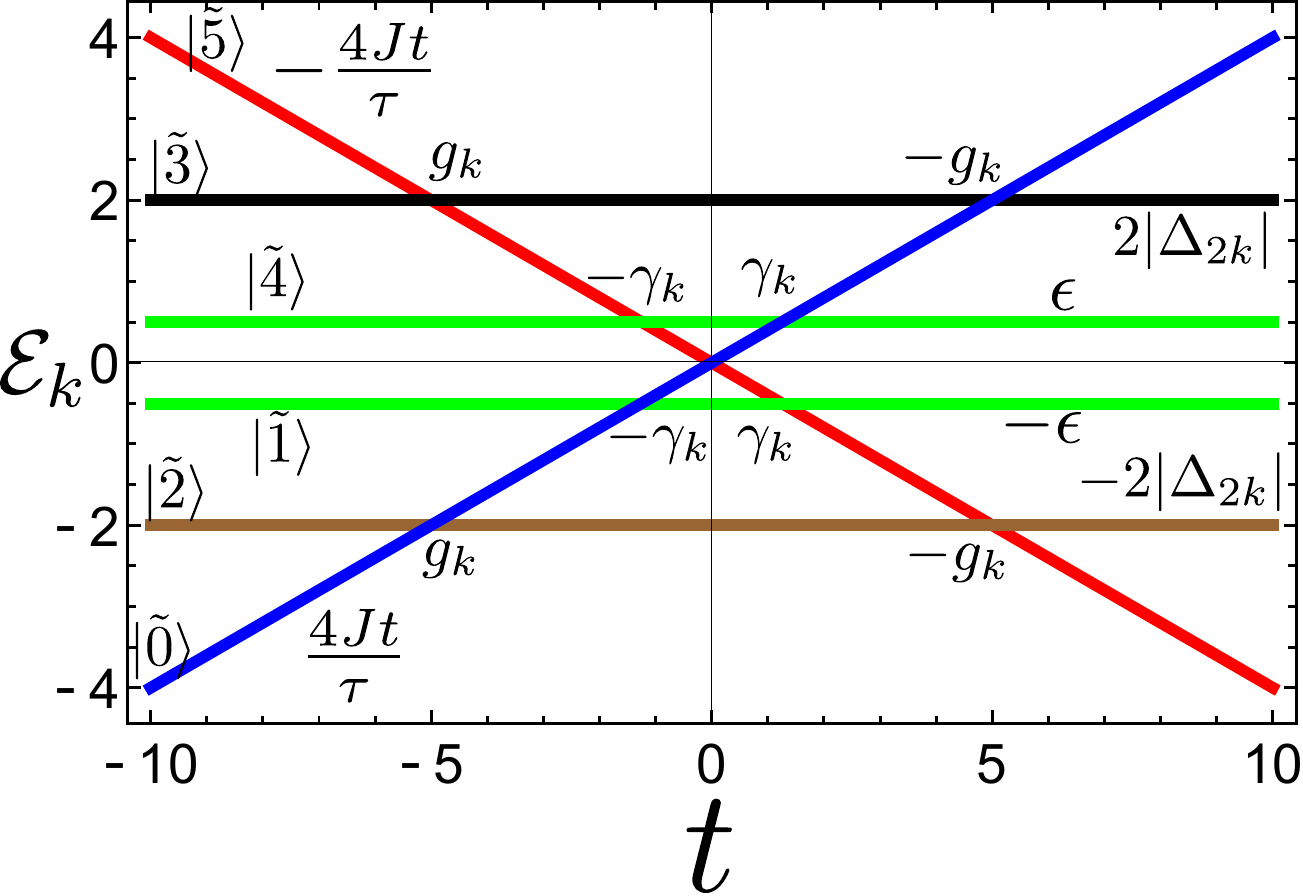}
    \caption{\textbf{Six-level problem, diabatic levels:} Plot of the diabatic energy levels corresponding to the six-level problem as a function of time. }
    \label{6stLZ}
\end{figure}
It is worthwhile to note that there is no direct coupling between the states $|0\rangle$ and $|5\rangle$. At each crossing point, the crossing involves a horizontal level (with zero slope) and a state with definite non-zero slope $\pm 4J/\tau$. Therefore, the denominator in the corresponding two-level L-Z transition amplitude is always $4J/\tau$ for all the available independent two-level L-Z transitions. There are only two types of the transition amplitudes for diabatic passage through any crossing point, viz., $u_{g} = \exp{\left( -\frac{\pi |g|^2 J \tau}{4 J^2} \right) }$, and $u_{\gamma} = \exp{\left( -\frac{\pi |\gamma|^2 J \tau}{4 J^2} \right) }$ whereas, the transition amplitude for the adiabatic passage through the crossing points are $i\xi \sqrt{1 - u_{g}^{2}}$, and $i\xi \sqrt{1 - u_{\gamma}^{2}}$, with $\xi =\pm 1$ as mentioned earlier.

 Let us start by finding all the amplitudes corresponding to the transitions from state $|0\rangle$ to all other states, viz., 
\begin{eqnarray}
 |\tilde{0}\rangle \rightarrow |\tilde{0}\rangle & : & S_{00} = (u_g u_{\gamma})^{2} e^{i \phi_{00}}, \nonumber \\
 |\tilde{0}\rangle \rightarrow |\tilde{2}\rangle & : & S_{02}, \nonumber \\
 |\tilde{0}\rangle \rightarrow |\tilde{1}\rangle & : & S_{01} = u_{\gamma} u_g (-i) \sqrt{1- u_{\gamma}^{2}} e^{i \phi_{02}},  \nonumber \\
 |\tilde{0}\rangle \rightarrow |\tilde{4}\rangle & : & S_{04} = u_{\gamma} u_g (i) \sqrt{1- u_{\gamma}^{2}} e^{i \phi_{03}}, \nonumber \\
 |\tilde{0}\rangle \rightarrow |\tilde{3}\rangle & : & S_{03} =  u_{\gamma}^{2} u_g (-i) \sqrt{1- u_{g}^{2}} e^{i \phi_{04}}, \nonumber \\
 |\tilde{0}\rangle \rightarrow |\tilde{5}\rangle & : & S_{05},
\end{eqnarray}

where $\phi_{0i}$'s are the dynamic phases associated with the paths. In the above,
\begin{eqnarray}
 S_{02} &=& \left[ (i)u_{g} \sqrt{1- u_{g}^{2}} e^{i \phi_{01}^{(1)}} + u_{g} \sqrt{1- u_{g}^{2}} (i)^3 (1-u_{\gamma}^{2}) e^{i \phi_{01}^{(2)}} \right], \nonumber \\
S_{05} &=& \left[ (-i)(i) (1- u_{g}^{2}) e^{i \phi_{05}^{(1)}} + u_{g}^{2} (-i)(i) (1-u_{\gamma}^{2}) e^{i \phi_{05}^{(2)}} \right]. \nonumber \\
\end{eqnarray}
In the above equations, the dynamic phases associated with the paths are given by,

\begin{eqnarray}
 & &\phi_{02}^{(1)} = \int_{-\infty}^{-T_2} \frac{4J t}{\tau} \; dt + \int_{-T_2}^{T_2} (- 2 |\Delta_{2k}|) \; dt  
 \nonumber \\
 & & +\int_{T_2}^{\infty} (- 2 |\Delta_{2k}|) \; dt, \nonumber \\
 & &\phi_{02}^{(2)} = \int_{-\infty}^{-T_1} \frac{4J t}{\tau} \; dt + \int_{-T_1}^{T_1} (- \epsilon) \; dt+ \nonumber \\
 & & \int_{T_1}^{T_2} \left(- \frac{4J t}{\tau} \right) \; dt  + \int_{T_2}^{\infty} (- 2 |\Delta_{2k}|) \; dt, \\ \nonumber \\
 & & \text{implying, in the limit } \epsilon \rightarrow 0,  \nonumber \\
 & &\Delta \phi_{02}  = \phi_{02}^{(1)} -\phi_{02}^{(2)} = -4 |\Delta_{2k}| T_2,
 \end{eqnarray}

 \begin{eqnarray}
 & & \phi_{05}^{(1)} = \int_{-\infty}^{-T_2} \frac{4J t}{\tau} \; dt + \int_{-T_2}^{T_2} (- 2 |\Delta_{2k}|) \; dt \nonumber \\  & &  + \int_{T_2}^{\infty} \left(- \frac{4J t}{\tau} \right) \; dt, \nonumber \\
 & &\phi_{05}^{(2)} = \left(\int_{-\infty}^{-T_2}+\int_{-T_2}^{-T_1}\right) \frac{4J t}{\tau} \; dt + \int_{-T_1}^{T_1} (- \epsilon) \; dt \nonumber \\ & & +  \left(\int_{T_1}^{T_2} + \int_{T_2}^{\infty}\right) \left(- \frac{4J t}{\tau} \right) \; dt, \\ \nonumber \\ 
 & & \text{implying, in the limit } \epsilon \rightarrow 0, \nonumber \\
 & &\Delta \phi_{05} = \phi_{05}^{(1)} -\phi_{05}^{(2)} = -4 |\Delta_{2k}| T_2.
\end{eqnarray}
It is worthwhile to note that once again (as was the case for four-state model) the dynamic phase difference is equal to the area under the curve. The transition probabilities are given by, $P_{00} = |S_{00}|^2 = (u_{g} u_{\gamma})^4$, this is the quantity we have defined as defect probability in  the equation \eqref{P00}. Similarly, $P_{01} = |S_{01}|^2 = P_{04} = |S_{04}|^2 = (u_{g} u_{\gamma})^2 (1-u_{\gamma}^{2})$, (note that they are equally probable) $P_{03} = |S_{03}|^2 = (u_{\gamma})^4 u_{g}^{2} (1-u_{g}^2)$, and
\begin{eqnarray}
 P_{02} &=& S_{01}^{*} S_{01} \nonumber \\ &=& u_{g}^{2} (1-u_{g}^{2}) [1 + (1-u_{\gamma}^{2})^{2} - 2 (1-u_{\gamma}^{2}) \sin (\Delta \phi_{01}) ], \nonumber \\
 P_{05} &=& S_{05}^{*} S_{05} \nonumber \\ &=& (1-u_{g}^{2})^2 + [u_{g}^{2} (1 - u_{\gamma}^2)]^2 + 2 (1-u_{\gamma}^{2}) u_{g}^{2} (1 - u_{g}^2) \nonumber \\ & & \times  \cos (\Delta \phi_{05}). 
\end{eqnarray}
Although we have defined $P_{00}$ only as the defect probability, the other probabilities except $P_{05}$ are still representing probabilities of the system going into different adiabatic states other than the adiabatic ground state $|5\rangle$ at $t = \infty$.  As a whole, the total probability must be one, i.e., $(P_{00}+ P_{01}+P_{02}+P_{03}+P_{04}+P_{05}) = 1$. Furthermore, it is easy to determine the relation between the time $T_2$ and the other parameters of the system to be $T_2 = \frac{|\Delta_{2k}|}{2J^2} (J\tau)$, from the fact that the slope of the diabatic energy curve, $\frac{2|\Delta_{2k}|}{T_2} = \frac{4J}{\tau}$. In the limit of $J\tau >> 1$, we expect the dynamic phase difference to be large enough and can be considered as a random phase.  

In the above, analysis we have considered all the transitions from the state $|\tilde{0}\rangle$ since it is the ground state of our 3D-Kitaev model at $t=-\infty$. One can follow the methodology explained above to get the entire transition probability matrix as has been done for 4-state representation of the model however, this is unnecessary for our purpose.

\section{Fitting of numerically obtained defect density }\label{ta3}
\begin{figure}
    \centering
    \includegraphics[scale=0.50]{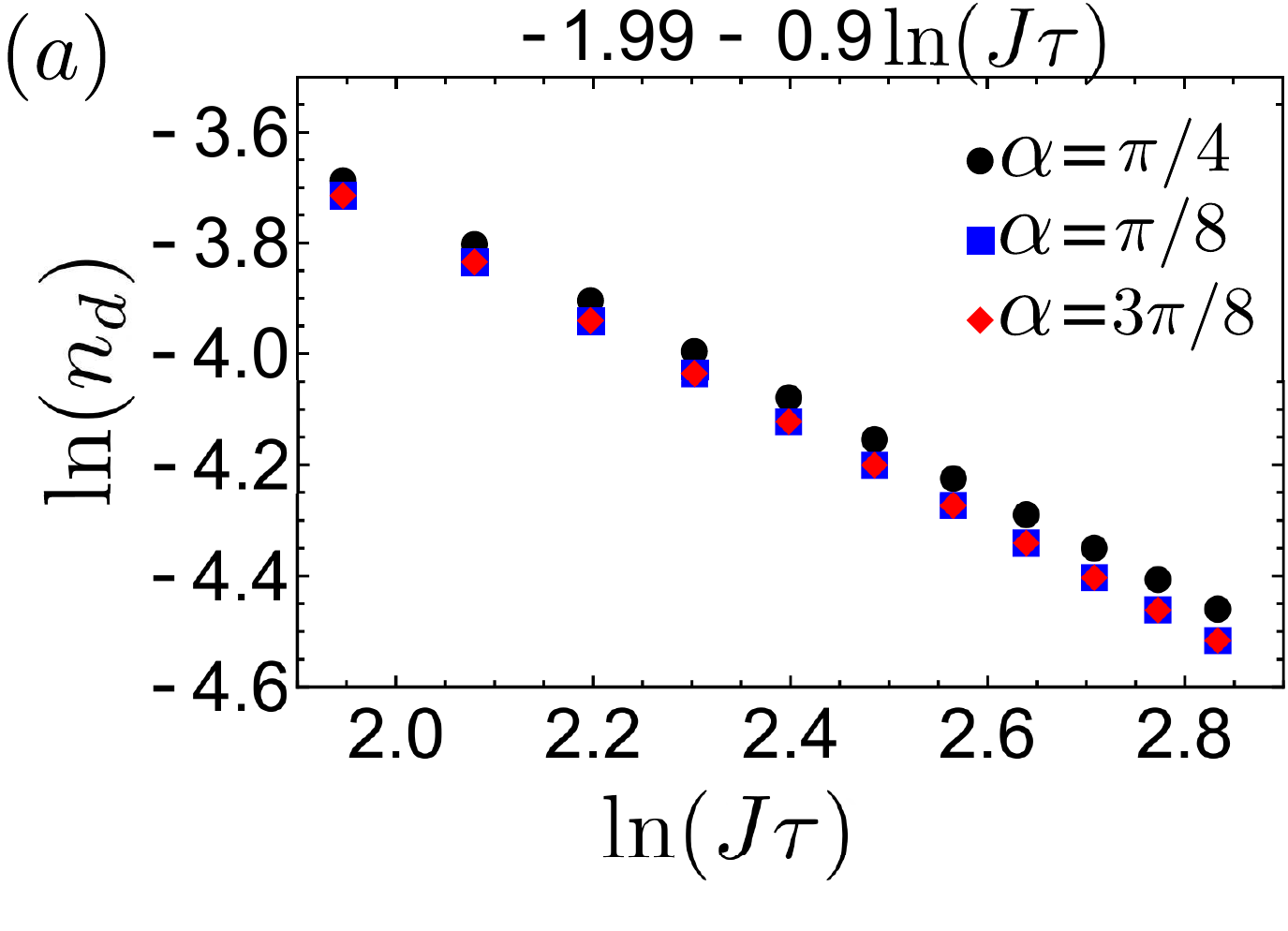}
    \includegraphics[scale=0.50]{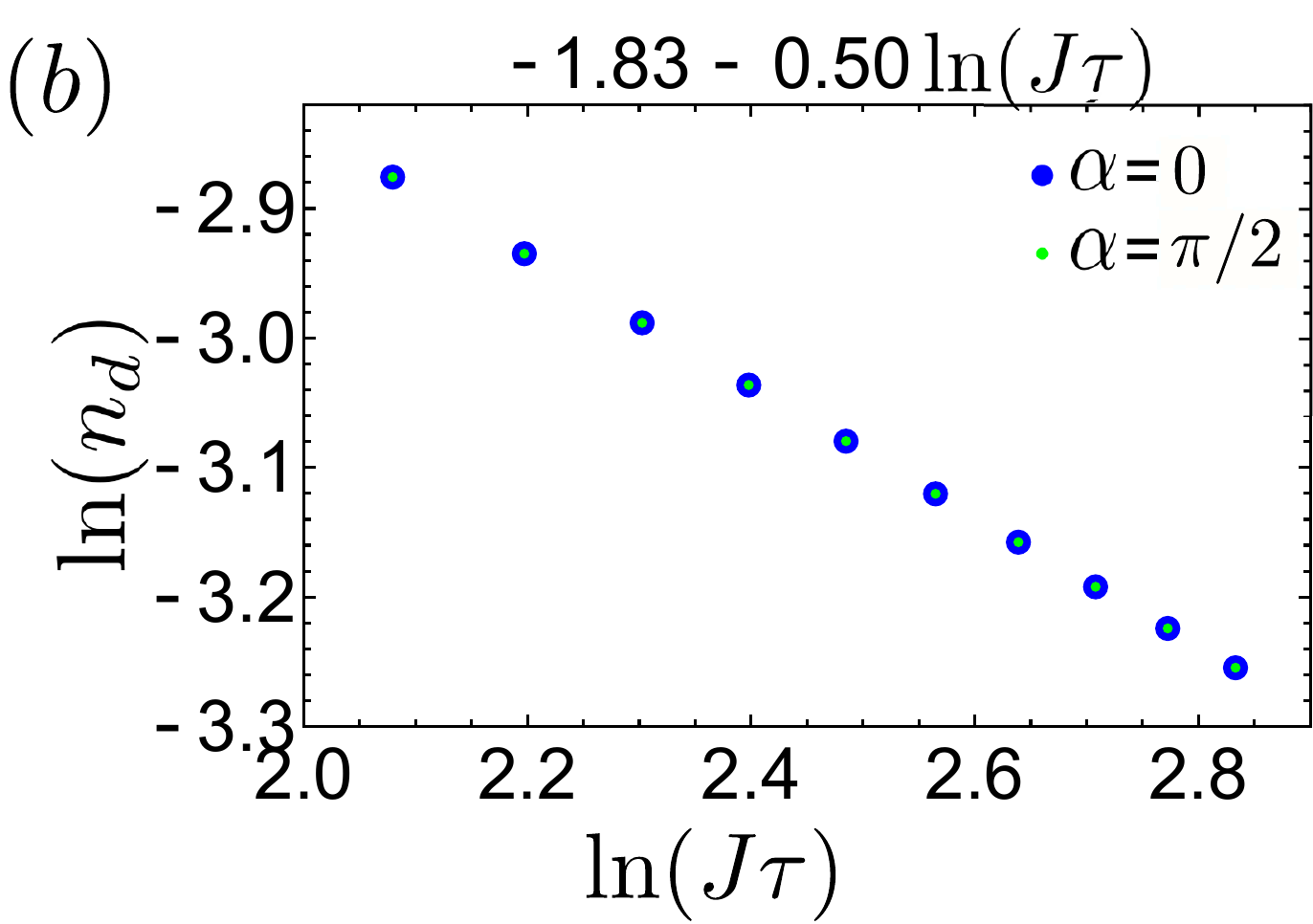}
    \caption{\textbf{Log-log plot of defect density as a function of $J\tau$}: (a) Plot of $\ln(n_d)$ vs $\ln(J\tau)$ in the $J\tau >> 1$ limit (in this case, $J\tau > 10$) for the numerically obtained defect density. Black circle, blue square and red diamond correspond to $\alpha = \pi /4,~  \pi /8$, and  $3 \pi /8$ respectively. (b) Plot of $\ln(n_d)$ vs $\ln(J\tau)$ in the $J\tau >> 1$ limit (in this case, $J\tau > 10$) for the numerically obtained defect density. Blue circle corresponds to $\alpha = 0$ and green circle corresponds to $\alpha = \pi /2$. The expressions at the top of both the plots represents $\ln (n_d) = \ln \left( \frac{A}{(J\tau)^{b}} \right) = \ln A - b \ln (J\tau)$, where $b$ determines the exponent corresponding to the scaling of the defect density.}
    \label{fit}
\end{figure}

In the following we have plotted the $\ln(n_d)$ corresponding to the six-level representation as a function of $\ln(J\tau)$ for different values of $\alpha$. We have found that when $\alpha = 0$ and $\pi / 2$ the exponent of the defect density is `$1/2$' which is as expected since in these two limits the 3D Kitaev model become disconnected Kitaev chains exhibiting a scaling of $\tau^{-1/2}$. For other values of $\alpha$ the defect density scales as $\tau^{-1}$ according to ICA and as shown in the Fig. \ref{fit} (a) the numerically obtained scaling is $\tau^{-0.9}$ which matches with the ICA value (within the numerical accuracy). We have checked that the four-level problem exhibit exactly the same quantitative behavior with respect to the scaling of the defect density. However, we have not provided the corresponding plots to avoid redundancy.
\section{Derivation of the correlation function expression corresponding to the equation \eqref{cacbk}}\label{tbc}

The correlation function as defined in \eqref{cacb} can be expressed in terms of the original $\psi (\mathbf{R})$ and $\phi(\mathbf{R})$ fermions in real space as follows,
\begin{eqnarray}
&& i c_{1a}(\mathbf{R}) c_{1b}(\mathbf{R}+\mathbf{r}) \nonumber \\
 =&& \frac{1}{4} \left( \psi(\mathbf{R}) + \psi^{\dagger}(\mathbf{R}) \right)  \left(\psi(\mathbf{R}+\mathbf{r})- \psi^{\dagger}(\mathbf{R}+\mathbf{r})  \right) \\
=&& \frac{1}{4}\Big(  \psi(\mathbf{R}) \psi(\mathbf{R}+\mathbf{r}) -   \psi(\mathbf{R}) \psi^{\dagger}(\mathbf{R}+\mathbf{r}) \nonumber \\
&& +   \psi^{\dagger}(\mathbf{R})  \psi(\mathbf{R}+\mathbf{r}) -  \psi^{\dagger}(\mathbf{R}) \psi^{\dagger}(\mathbf{R}+\mathbf{r})  \Big) 
\end{eqnarray}
Now we  use the Fourier transformation $\psi(\mathbf{R})= \frac{1}{\sqrt{N}} \int d^ k e^{i \mathbf{k} \cdot \mathbf{R}} \psi_k$. We notice that contributions from the 1st and the last term are zero, and contributions from the second and the third are conjugate to each other. With the above expressions we use the definitions in \eqref{ukpk} to re-express correlation function in terms of $\alpha_{ik}$ and using orthogonality of states we arrive at \eqref{cacbk}.
\end{appendix}
\bibliography{Kitaev_defect}
\end{document}